\documentclass{emulateapj}
\usepackage{apjfonts}
\usepackage{epsfig}

\journalinfo{The Astrophysical Journal, in press}
\slugcomment{Received 2009 June 3; accepted 2009 August 18}

\shortauthors{CAMILO ET AL.}
\shorttitle{TWO RADIO-EMITTING GAMMA-RAY PULSARS}

\begin{document}

\def\chandra{{\em Chandra}}
\def\fermi{{\em Fermi}}
\def\rosat{{\em ROSAT}}
\def\swift{{\em Swift}}
\def\xmm{{\em XMM}}
\def\ergs{\,erg\,s$^{-1}$}
\def\pccc{\,pc\,cm$^{-3}$}

\def\cyg{Cyg~OB2}
\def\cygf{Cyg~OB2~4}
\def\mt{MT91~213}
\def\cxou{CXOU~J203213.5+412711}
\def\egreta{3EG~J1741--2050}
\def\egretb{3EG~J2033+4118}
\def\egr{EGR~J2033+4117}
\def\tev{TeV~J2032+4130}
\def\psra{PSR~J1741--2054}
\def\psrb{PSR~J2032+4127}
\def\swifta{Swift~J174157.6--205411}

\title{Radio detection of LAT PSRs~J1741--2054 and J2032+4127: no longer
just gamma-ray pulsars}

\author{F.~Camilo\altaffilmark{1},
  P.~S.~Ray\altaffilmark{2},
  S.~M.~Ransom\altaffilmark{3},
  M.~Burgay\altaffilmark{4},
  T.~J.~Johnson\altaffilmark{5,6},
  M.~Kerr\altaffilmark{7},
  E.~V.~Gotthelf\altaffilmark{1},
  J.~P.~Halpern\altaffilmark{1},
  J.~Reynolds\altaffilmark{8,9},
  R.~W.~Romani\altaffilmark{10},
  P.~Demorest\altaffilmark{3},
  S.~Johnston\altaffilmark{8},
  W.~van~Straten\altaffilmark{11},
  P.~M.~Saz~Parkinson\altaffilmark{12},
  M.~Ziegler\altaffilmark{12},
  M.~Dormody\altaffilmark{12},
  D.~J.~Thompson\altaffilmark{5},
  D.~A.~Smith\altaffilmark{13,14},
  A.~K.~Harding\altaffilmark{5},
  A.~A.~Abdo\altaffilmark{2,15},
  F.~Crawford\altaffilmark{16},
  P.~C.~C.~Freire\altaffilmark{17},
  M.~Keith\altaffilmark{8},
  M.~Kramer\altaffilmark{18,19},
  M.~S.~E.~Roberts\altaffilmark{20},
  P.~Weltevrede\altaffilmark{8},
  and K.~S.~Wood\altaffilmark{2}
}

\altaffiltext{1}{Columbia Astrophysics Laboratory, Columbia University,
  New York, NY~10027, USA}
\altaffiltext{2}{Space Science Division, Naval Research Laboratory,
  Washington, DC 20375, USA}
\altaffiltext{3}{National Radio Astronomy Observatory, Charlottesville, 
  VA~22903, USA}
\altaffiltext{4}{INAF --- Osservatorio Astronomico di Cagliari, 09012
  Capoterra, Italy}
\altaffiltext{5}{NASA Goddard Space Flight Center, Greenbelt, MD~20771,
  USA}
\altaffiltext{6}{University of Maryland, College Park, MD~20742, USA}
\altaffiltext{7}{Department of Physics, University of Washington, Seattle,
  WA~98195, USA}
\altaffiltext{8}{Australia Telescope National Facility, CSIRO, Epping,
  NSW~1710, Australia}
\altaffiltext{9}{CSIRO Parkes Observatory, Officer-in-Charge for life,
  Parkes, NSW~2870, Australia}
\altaffiltext{10}{Department of Physics, Stanford University, Stanford,
  CA~94305, USA}
\altaffiltext{11}{Centre for Astrophysics and Supercomputing, Swinburne
  University of Technology, Hawthorn, VIC~3122, Australia}
\altaffiltext{12}{Santa Cruz Institute for Particle Physics, Department
  of Physics and Department of Astronomy and Astrophysics, University
  of California at Santa Cruz, CA~95064, USA}
\altaffiltext{13}{CNRS/IN2P3, Centre d'\'Etudes Nucl\'eaires de Bordeaux-Gradignan, Gradignan, 33175, France}
\altaffiltext{14}{Universit\'e de Bordeaux, Centre d'\'Etudes Nucl\'eaires
  de Bordeaux-Gradignan, Gradignan, 33175, France}
\altaffiltext{15}{NRC Research Associate}
\altaffiltext{16}{Department of Physics and Astronomy, Franklin and
  Marshall College, Lancaster, PA~17604, USA}
\altaffiltext{17}{NAIC, Arecibo Observatory, Arecibo, PR~00612, USA}
\altaffiltext{18}{MPIfR, 53121~Bonn, Germany}
\altaffiltext{19}{Jodrell Bank Centre for Astrophysics, University of
  Manchester, Manchester M13~9PL, UK}
\altaffiltext{20}{Eureka Scientific, Inc., Oakland, CA~94602, USA}

\begin{abstract}
Sixteen pulsars have been discovered so far in blind searches of photons
collected with the Large Area Telescope on the {\em Fermi Gamma-ray
Space Telescope}.  We here report the discovery of radio pulsations
from two of them.  \psra, with period $P=413$\,ms, was detected in
archival Parkes telescope data and subsequently has been detected at the
Green Bank Telescope (GBT).  Its received flux varies greatly due to
interstellar scintillation and it has a very small dispersion measure
of $\mbox{DM}=4.7$\pccc, implying a distance of $\approx 0.4$\,kpc and
possibly the smallest luminosity of any known radio pulsar.  At this
distance, for isotropic emission, its gamma-ray luminosity above
0.1\,GeV corresponds to 25\% of the spin-down luminosity of $\dot E =
9.4\times10^{33}$\ergs.  The gamma-ray profile occupies 1/3 of pulse phase
and has three closely-spaced peaks with the first peak lagging the radio
pulse by $\delta = 0.29\,P$.  We have also identified a soft \swift\
source that is the likely X-ray counterpart.  In many respects \psra\
resembles the Geminga pulsar.  The second source, \psrb, was detected
at the GBT.  It has $P=143$\,ms, and its $\mbox{DM}=115$\pccc\ suggests
a distance of $\approx 3.6$\,kpc, but we consider it likely that it
is located within the \cyg\ stellar association at half that distance.
The radio emission is nearly 100\% linearly polarized, and the main radio
peak precedes by $\delta = 0.15\,P$ the first of two narrow gamma-ray
peaks that are separated by $\Delta = 0.50\,P$.  The second peak has a
harder spectrum than the first one, following a trend observed in young
gamma-ray pulsars.  Faint, diffuse X-ray emission in a \chandra\ image is
possibly its pulsar wind nebula.  \psrb\ likely accounts for the EGRET
source \egretb, while its pulsar wind is responsible for the formerly
unidentified HEGRA source \tev.  \psrb\ is coincident in projection with
\mt, a Be star in \cyg, although apparently not a binary companion of it.

\end{abstract}


\section{Introduction} \label{sec:intro} 

Neutron stars hold vast stores of rotational energy.  Magnetic braking
in such stars generates spin-down luminosities of $\dot E \approx
10^{28-38}$\ergs.  A minute portion of this emerges in collimated
coherent radio beams through which most known rotation-powered pulsars
are detected.  Much in pulsar electrodynamics remains obscure, including
processes and locations of particle acceleration in the magnetosphere.
In contrast to radio emission, pulsed gamma-ray luminosity can be a
substantial fraction of $\dot E$, and its study therefore holds promise
for significant advance in understanding magnetized rotating neutron
stars.

Until recently, only six pulsars were confirmed gamma-ray emitters above
0.1\,GeV \citep{tho04b}.  Their detection in the 1990s with the EGRET
experiment on the {\em Compton Gamma-Ray Observatory} used rotational
ephemerides to fold the sparse gamma-ray photons at a known period.  The
ephemerides were obtained from radio observations, with the exception of
Geminga, which is not a detected radio source.  Given these limitations,
it was not possible, for instance, to establish the fraction of gamma-ray
pulsars that do not emit radio beams detectable at the Earth, i.e., that
are ``radio quiet''.  According to predictions of outer magnetosphere
models \citep[e.g.,][]{ry95}, gamma-ray beams are broader than radio
beams and thus are potentially detectable from a larger solid angle.
Measuring the fraction of radio-quiet gamma-ray pulsars therefore can
provide information on beam shapes and emission regions, and constrain
emission mechanisms \citep[e.g.,][]{hgg07}.

The {\em Fermi Gamma-ray Space Telescope}, operational since mid-2008,
is being used to make a torrent of pulsar discoveries with its Large
Area Telescope (LAT), which has substantially greater effective area,
field of view, and angular resolution than EGRET.  Using radio-derived
ephemerides, at least 14 pulsars have been detected for the first time by
LAT, of which eight are millisecond pulsars \citep{abdo5,abdo1,abdo10}.
In addition, at least 16 pulsars have been discovered in periodicity
searches of gamma-ray photons \citep{abdo2}.  While a few of these
had been searched previously in very sensitive radio observations, and
therefore may be considered radio quiet \citep{bwa+04,hgc+04,hcg07},
most new LAT pulsars should be searched for radio counterparts before
their true character can be determined.  A radio detection yields an
estimate of distance and therefore of luminosity, allowing for a more
comprehensive study of these energetic neutron stars.  Here we report
the detection of radio pulsations, as well as additional gamma-ray and
X-ray analysis, from two \fermi\ LAT pulsars.

\section{Observations and Results} \label{sec:obs} 

Nine of the new LAT pulsars are located in areas previously surveyed with
the ATNF Parkes radio telescope.  Three had been targeted in sensitive
but unsuccessful searches \citep{rhr+02}.  We searched anew these
data, as well as archival 1.4\,GHz data for the other six LAT pulsars
\citep[J0633+0632, J1459--60, J1732--31, J1741--2054, J1813--1246,
and J1907+06;][]{abdo2}.  The 95\%~C.L. error radius for these LAT
sources spanned $4\farcm6$--$8\farcm4$.  With a beam radius of $7'$,
there were in general multiple relevant Parkes pointings to inspect.
We searched each such data set in dispersion measure (DM) with PRESTO
\citep{ran01,rem02}, folding time samples from the 35 minute integrations
modulo the pulsar period predicted from each LAT ephemeris.  In the case
of \psra, we detected pulsations.

\begin{figure}[t]
\begin{center}
\includegraphics[angle=0,scale=0.75]{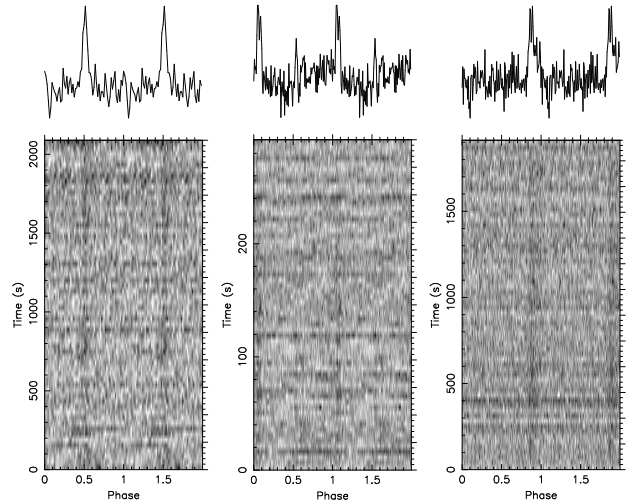}
\caption{\label{fig:psrar}
Radio detections of \psra: Parkes at 1.4\,GHz ({\em left}); GBT at 2\,GHz
({\em center}), and at 350\,MHz ({\em right}).  In each panel the pulse
profile is shown repeated in phase, as a function of time in the grey
scale, and summed at the top.  The relative pulse phases between panels
are arbitrary.
}
\end{center} 
\end{figure}

\subsection{\psra } \label{sec:psra}

\subsubsection{Radio measurements} \label{sec:radioa}

The 1.4\,GHz detection of the $P=413$\,ms \psra\ (left panel of
Figure~\ref{fig:psrar}) is from data collected across a bandwidth
of 288\,MHz in the Parkes multibeam survey of the Galactic plane
\citep[e.g.,][]{mlc+01} on 2000 November 24.  This location ($\mbox{R.A.}
= 17^{\rm h}41^{\rm m}51\fs3$, $\mbox{decl.} = -21\arcdeg01'10''$;
all positions here refer to the J2000.0 equinox) was $7\farcm6$ away
from the nominal LAT position, which had an uncertainty of $8\farcm4$
\citep{abdo1}.

In order to further constrain its position, we attempted
several observations of the pulsar with the NRAO Green Bank
Telescope (GBT) at 2\,GHz, for which the beam radius is $3'$, using
GUPPI\footnote{https://wikio.nrao.edu/bin/view/CICADA/GUPPiUsersGuide}
with approximately 700\,MHz of usable bandwidth.  We did this at six
epochs, each time on a grid of pointings surrounding the nominal Parkes
position.  We only detected the pulsar on 2009 March 7, at $\mbox{R.A.}
= 17^{\rm h}41^{\rm m}50\fs1$, $\mbox{decl.} = -20\arcdeg55'02''$,
$4\farcm6$ from the LAT position.  The received flux varied greatly
during the 5 minute observation (middle panel of Figure~\ref{fig:psrar}),
possibly due to interstellar scintillation.

We obtained from the first two detections only an upper limit on DM.
We finally measured the dispersion of \psra\ with a 350\,MHz detection
on March 12 (right panel of Figure~\ref{fig:psrar}) using the GBT BCPM
\citep{bdz+97} with a bandwidth of 48\,MHz: $\mbox{DM} \approx 5$\pccc.
Such a low DM points to scintillation \citep[see, e.g.,][]{lk05}
as an explanation for the several non-detections (which further
include at Parkes two observations at 1.4\,GHz and one at 3\,GHz,
and at GBT one 0.8\,GHz observation).  For this DM and location
at $(l,b) = (6\fdg42,+4\fdg90)$, the \citet{cl02} electron density
model yields a pulsar distance of $d = 0.4$\,kpc, with significant
uncertainty.  Hereafter, we parameterize this distance by $d_{0.4} =
d/(0.4\,\mbox{kpc})$.

We have estimated the observed flux densities $S_\nu$ for the three radio
detections: $S_{0.35} \approx 1.33$\,mJy, $S_{1.4} \approx 0.16$\,mJy,
and $S_{2} \approx 0.09$\,mJy.  However, because the detection positions
at the two highest frequencies are slightly offset from the true position
(e.g., Figure~\ref{fig:swift}), these values somewhat underestimate the
flux on those days.  Conversely, because of scintillation, the average
flux is likely substantially lower than the position-corrected values
on those days.  Thus we cannot infer a spectral index.

A comparison of the barycentric pulsar periods determined from radio
observations taken more than 8 years apart implies a period derivative
of $\dot P = (2.7\pm1.1) \times10^{-14}$ (unless otherwise noted, all
uncertainties here are given at the $1\,\sigma$ C.L.).

\subsubsection{Gamma-ray and radio timing and position} \label{sec:timinga}

\begin{figure}[t]
\begin{center}
\includegraphics[angle=0,scale=0.50]{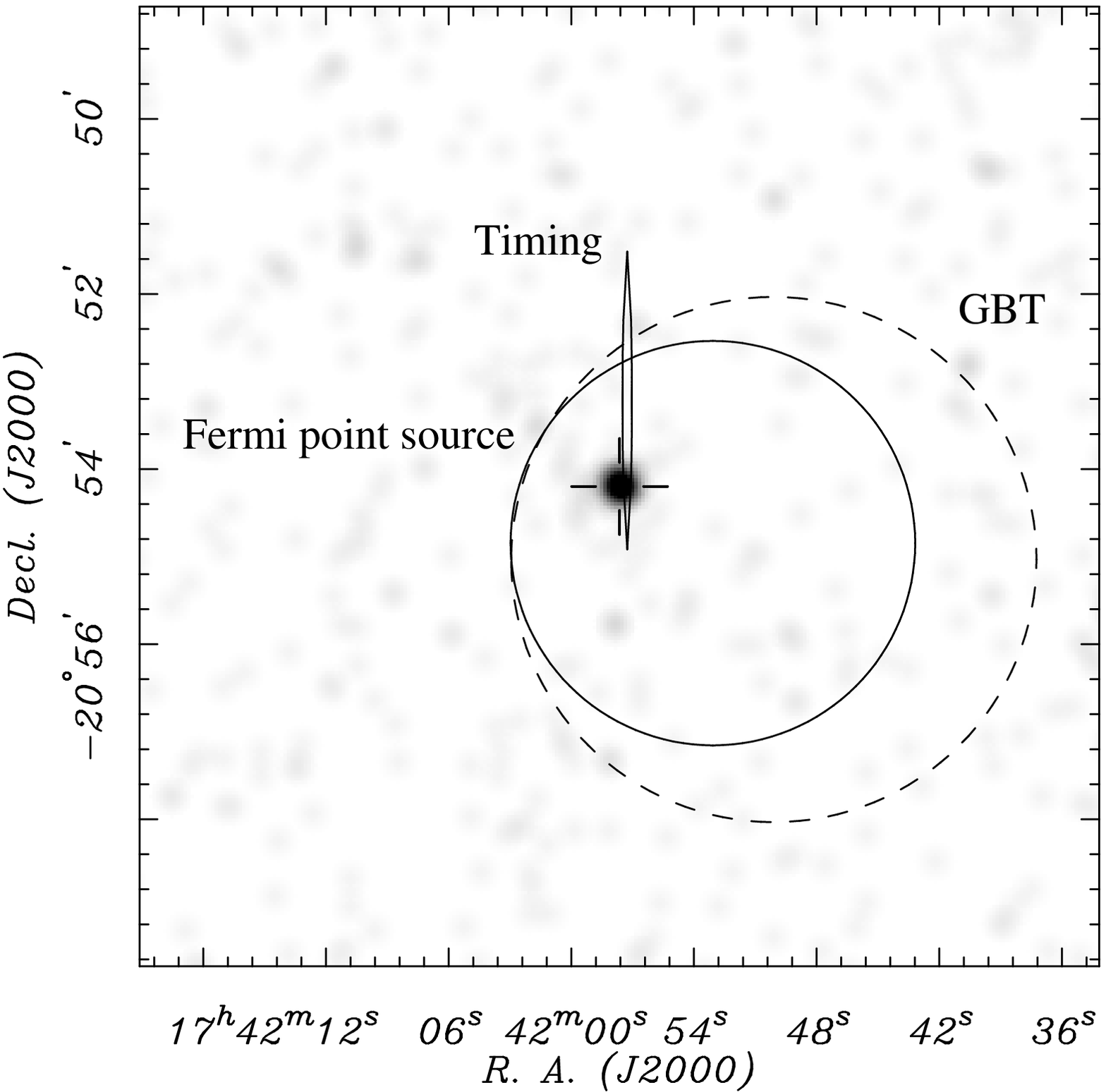}
\caption{\label{fig:swift}
X-ray field of \psra: portion of \swift\ 0.2--10\,keV image smoothed
with a $\sigma = 7''$ Gaussian.  The soft source \swifta, with
69 photons within an extraction radius of $0\farcm5$, is the only
plausible counterpart in this 4.4\,ks observation for \psra, whose
$3\,\sigma$ timing position ({\em ellipse}), 95\%~C.L. gamma-ray point
source ({\em solid circle}), and FWHM radio ({\em dashed circle})
positional uncertainties are indicated (see \S\S~\ref{sec:timinga}
and \ref{sec:xraya}).
}
\end{center}
\end{figure}

Using LAT data spanning one year, we have extracted 17 gamma-ray
times-of-arrival (TOAs) for \psra\ that we use to obtain a timing solution
with TEMPO\footnote{http://www.atnf.csiro.au/research/pulsar/tempo}.
While the fitted position in R.A.\ has an uncertainty of only $1''$,
that in decl.\ is uncertain by $0\farcm6$ (see Figure~\ref{fig:swift}),
owing to the low ecliptic latitude of the pulsar.  In addition, a \swift\
observation reveals a probable X-ray counterpart at $\mbox{R.A.} =
17^{\rm h}41^{\rm m}57\fs6$, $\mbox{decl.} = -20\arcdeg54'11''$, with
uncertainty of about $4''$.  The \swift\ source position is $5''$ in R.A.\
and $58''$ in decl.\ from the timing position, consistent within the
combined $2\,\sigma$ uncertainties.  Because of the still significant
positional uncertainty (which should be essentially eliminated with a
planned \chandra\ observation), we cannot rule out the presence of a
small amount of timing noise.  In any case, the timing solution already
yields an accurate $\dot P = 1.69\times10^{-14}$.  In turn, $\tau_c =
P/(2\dot P) = 0.39$\,Myr and $\dot E = 9.4\times 10^{33}$\ergs.

\begin{figure}[t]
\begin{center}
\includegraphics[angle=0,scale=0.78]{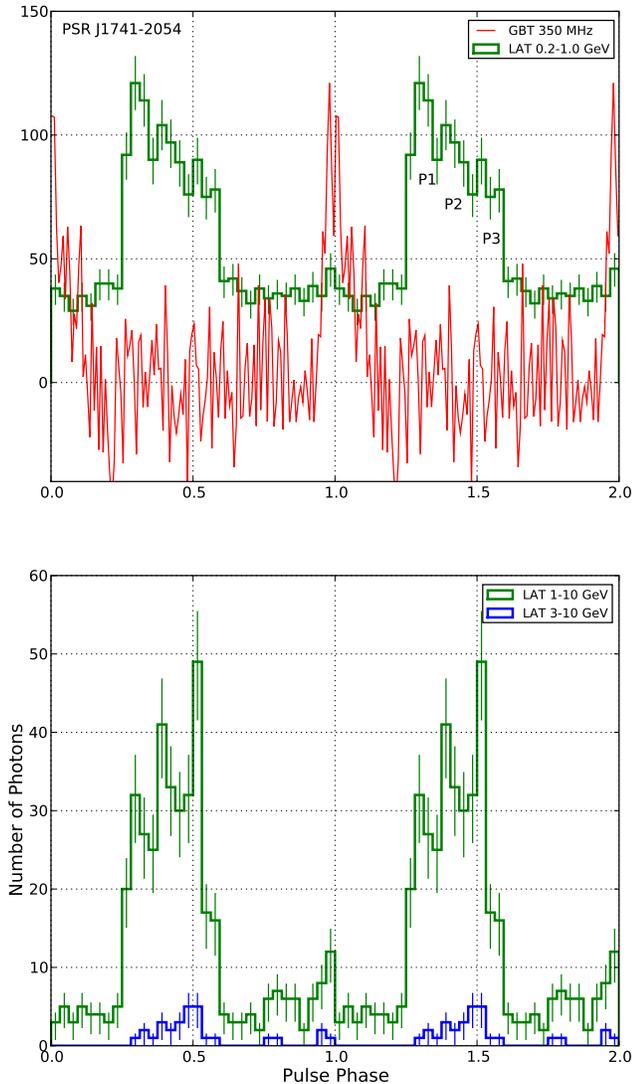}
\caption{\label{fig:psrag}
Phase-aligned \fermi\ and GBT pulse profiles of \psra.  In the upper
panel, the radio profile is displayed with an arbitrary intensity scale
along with LAT counts in the 0.2--1.0\,GeV band.  The bottom panel shows
the higher-energy LAT counts and a comparison between the two panels shows
clear evolution of the peak structure (P1 weakening, P3 strengthening)
with energy.  At the highest energies ($>3$\,GeV), P3 dominates.
The displayed gamma-ray and radio profiles have, respectively, 32 and
128 bins per period.  Two full rotations are shown.
}
\end{center} 
\end{figure}

We determined the position for the \psra\ LAT point source with
the binned likelihood application {\em ptlike}, by selecting photons
above 0.125\,GeV from the light curve peak, with phases $0.24 < \phi <
0.55$ (Figure~\ref{fig:psrag}; this phase range subtends roughly the
half-maximum level of pulsed emission, offering the best signal-to-noise
ratio for localization), and varying the position to maximize the
point source significance.  The resulting position is $\mbox{R.A.} =
17^{\rm h}41^{\rm m}53\fs1$, $\mbox{decl.} = -20\arcdeg54'51''$, with
a 95\%~C.L. error radius of $2\farcm3$.  This is only $1\farcm2$ away
from the \swift\ source.  The best radio position is also consistent with
the LAT timing and \swift\ source positions (see Figure~\ref{fig:swift}
for an illustration of the various positional determinations).

Using the 350\,MHz and 2\,GHz TOAs in a separate fit, we also
obtain a higher-precision measurement of dispersion, $\mbox{DM} =
(4.7\pm0.1)$\pccc.  This allows the determination of the phase
offset between the radio and gamma-ray pulses separately for both
recent radio detections.  The results are identical, and we show in
Figure~\ref{fig:psrag} the phase relation for the 350\,MHz and gamma-ray
profiles.  The LAT profiles are constructed from ``diffuse class''
photons \citep{atwood} selected from a region with an energy-dependent
radius $0\fdg8\,(E/{\rm 1\,GeV})^{-0.75}$ (with maximum and minimum
radii of $0\fdg8$ and $0\fdg35$, respectively) around the best position
over the time range 2008 August 4 to 2009 July 4.  We used energies
from 0.2--10\,GeV.  The phase alignment was done using the {\em
gtpphase}\footnote{http://fermi.gsfc.nasa.gov/ssc/data/analysis} tool
to assign phases to each LAT photon according to the best timing model.
We found that the pulsed emission was well-described by a three-Gaussian
model and extracted the parameters with a maximum likelihood fit to
the unbinned profile photon phases.  We applied the model to the energy
bands shown in Figure~\ref{fig:psrag} (0.2--1.0\,GeV and 1--10\,GeV) as
well as to the full (0.2--10\,GeV) energy range.  We found no significant
difference in the widths and positions of the Gaussians, so we report here
the results from a fit to the full energy range.  The first gamma-ray peak
(``P1''), with phase $0.292 \pm 0.005$ and $\mbox{FWHM} = 0.06 \pm 0.01$,
determines the phase offset with the radio peak, $\delta = 0.29\pm0.02$.
The second gamma-ray peak (``P2'') occurs at $\phi = 0.384 \pm 0.015$
with $\mbox{FWHM} = 0.13 \pm 0.03$, while P3 is located at $\phi =
0.518 \pm 0.015$ with $\mbox{FWHM} = 0.11 \pm 0.02$.  Interpreting the
P3--P1 separation as the conventional separation in a two-peak profile
yields $\Delta = 0.226 \pm 0.016$, although it is unclear whether this
is meaningful for such a complex profile.

\subsubsection{Gamma-ray spectrum} \label{sec:gammaa}

To fit the gamma-ray spectrum of \psra, we have performed an unbinned
likelihood analysis with the application {\em gtlike}\footnotemark[23],
and an additional check using the independent {\em ptlike}.  Both
approaches are outlined in \citet{abdo3}, and here we use an updated
instrument response function, \verb#P6_v3_diff#\footnotemark[23],
that corrects a pileup effect identified in orbit.  The on-axis energy
resolution of LAT is (equivalent Gaussian $1\,\sigma$) 15\%--9\% in the
0.1--1\,GeV range, 8\% over 1--10\,GeV, and 8\%--18\% in the 10--300\,GeV
range \citep{atwood}.  We used \verb#gll_iem_v02#\footnotemark[23]
as the model for the Galactic diffuse background.  We selected
gamma-rays with energy $>0.1$\,GeV and with zenith angle $<105\arcdeg$,
extracted from within a radius of $15\arcdeg$ around the pulsar
over the time range 2008 August 4 to 2009 July 4.  We modeled all
significant point sources in the aperture with a power-law spectrum.
We found no evidence of point source emission off pulse ($0.68 < \phi <
1.18$), so to increase the signal-to-noise ratio, we considered only
photons with phase $0.18 < \phi < 0.68$.  The likelihood ratio test
indicates that an exponentially cut-off power law given by $dN/dE =
N_0\,(E/\mbox{1\,GeV})^{-\Gamma} \exp(-{E/E_c})$ is preferred over
a simple power-law model with $12\,\sigma$ significance.  We find a
photon index $\Gamma = 1.4\pm0.1\pm0.1$ and a cut-off energy $E_c =
(1.1\pm0.2\pm0.2)$\,GeV, giving a photon flux above 0.1\,GeV of
$(20\pm1\pm3)\times 10^{-8}$\,cm$^{-2}$\,s$^{-1}$ and an energy flux of
$F_{\gamma} = (12\pm1\pm2)\times 10^{-11}$\,erg\,cm$^{-2}$\,s$^{-1}$,
where the first (second) errors indicated are statistical (systematic).
The resulting gamma-ray luminosity, assuming effectively isotropic
emission ($f_{\Omega} = 1$), is $L_{\gamma} = 4\pi f_{\Omega}
F_{\gamma} d^2 = 2.4\times10^{33}\,d_{0.4}^2\,\mbox{erg\,s$^{-1}$} =
0.25\,d_{0.4}^2\,\dot E$.

The third peak has a significantly harder spectrum than P1: it is not
detectable at 0.1--0.3\,GeV, then becomes gradually more significant with
increasing energy, such that above 1\,GeV its peak count rate exceeds
that of P1.  The spectrum of P2 appears to be of intermediate hardness
between that of P3 and P1 (see Figure~\ref{fig:psrag}).  The highest
energy photon likely detected so far from this pulsar arrived at phase
0.491 (in P3) with energy 5.8\,GeV.

\subsubsection{Spectrum of likely X-ray counterpart} \label{sec:xraya}

The error circle of \psra\ was observed with the \swift\ X-ray
telescope on 2008 October 16, obtaining 4.4\,ks of exposure with
2.5\,s time sampling.  The soft source \swifta\ near the field center
is unresolved and provides 69 background-subtracted counts in the
0.3--5\,keV range.  There is no other detected X-ray source that is
compatible with the position of the pulsar (Figure~\ref{fig:swift}),
and the spectral character of this source (see below) is compatible
with our expectations for \psra.  We therefore believe that \swifta\
is the likely pulsar counterpart.

We see no evidence in \swifta\ for interstellar absorption.  If we fit
the source spectrum with a fixed $N_{\rm H}= 1.5 \times 10^{20}\,{\rm
cm^{-2}}$ (10 times the DM-determined free electron column), we obtain
a power-law index $\Gamma=2.5\pm0.4$ and $F_X (0.5-10\,{\rm keV}) =
6.4 \times 10^{-13}\,{\rm erg\,cm^{-2}\,s^{-1}}$.  Allowing $N_{\rm
H}$ to vary yields $N_{\rm H} < 3 \times 10^{21}\,{\rm cm^{-2}}$ and
a poorly determined $\Gamma=2.3\pm1.7$.  Thermal models give $kT =
(0.2\pm0.1)$\,keV but are not well constrained because of the unknown
absorption.

The observed X-ray flux likely represents a composite spectrum similar
to other relatively old and nearby gamma-ray pulsars such as Geminga and
PSR~B1055--52.  At an age comparable to the characteristic age of \psra,
0.4\,Myr, thermal emission from the full surface is expected to have a
relatively low $kT \approx 50-70$\,eV.  Such low temperatures alone do not
reproduce the \swift\ spectrum which has significant counts above 1\,keV.
Thus the observed X-rays likely have a significant synchrotron component
coming from secondary $e^{\pm}$ pairs that are gamma-ray annihilation
products.

\subsection{\psrb } \label{sec:psrb}

\subsubsection{Radio measurements} \label{sec:radiob}

\begin{figure}[t]
\begin{center}
\includegraphics[angle=0,scale=0.70]{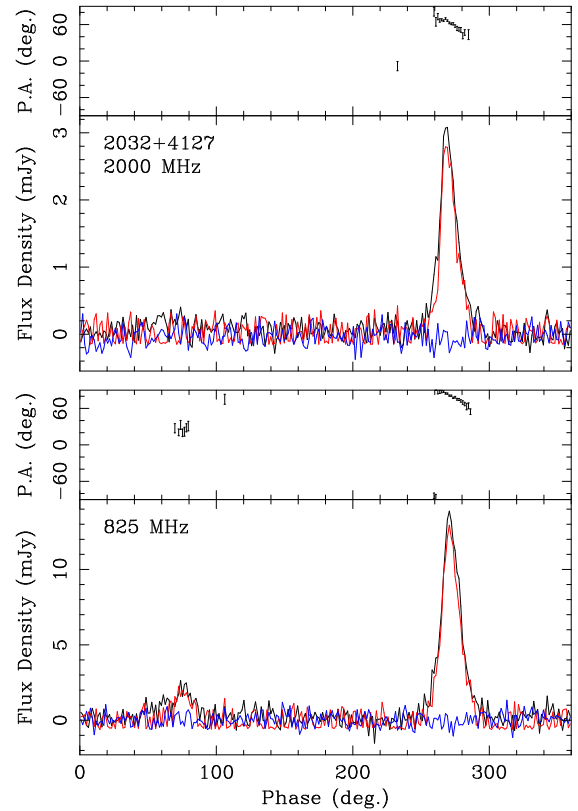}
\caption{\label{fig:psrbr}
Polarimetric profiles of \psrb\ based on GBT observations with GUPPI:
0.5\,hr at 2\,GHz ({\em top}) and 2.0\,hr at 0.8\,GHz ({\em bottom}).
The black traces corresponds to total intensity, while the red and blue
lines correspond, respectively, to linear and circular polarization.
The position angles in each upper panel have been converted to the pulsar
frame, using the measured $\mbox{RM} = (+215\pm1)$\,rad\,m$^{-2}$.
}
\end{center} 
\end{figure}

On 2009 January 5 we used the GBT to search for radio pulsations from
LAT \psrb.  We recorded data for 1\,hr from an 800\,MHz band centered
on 2\,GHz, using GUPPI to sample each of 1024 frequency channels every
0.16\,ms.  The telescope beam, with radius $3'$, was centered on the
LAT source at $\mbox{R.A.} = 20^{\rm h}32^{\rm m}13\fs9$, $\mbox{decl.}
= +41\arcdeg22'34''$ with error radius of $5\farcm1$ \citep{abdo1}.

We did a search in DM while folding the data modulo the 71.6\,ms period
of the original LAT ephemeris, and detected the pulsar.  Subsequently
we did a blind search of the radio data and established that its true
period is $P=143$\,ms.  Following some gridding observations, we obtained
an improved position: $\mbox{R.A.} = 20^{\rm h}32^{\rm m}14^{\rm s}$,
$\mbox{decl.} = +41\arcdeg27'00''$, with uncertainty radius of $0\farcm7$.
At this accurate position, the pulsar is clearly detectable at the GBT
in only 1 minute.

We have also obtained two calibrated polarimetric observations
\citep[analyzed with PSRCHIVE;][]{hvm04}, one each at 2\,GHz and 0.8\,GHz,
which show that the profile is nearly 100\% linearly polarized, with a
small interpulse detected at the lower frequency (Figure~\ref{fig:psrbr}).
The period-averaged flux density at 2\,GHz is $S_2 = 0.12$\,mJy,
and $S_{0.8} = 0.65$\,mJy.  Each of these measurements has fractional
uncertainty of about 20\%, including an allowance for the small degree
of interstellar scintillation inferred from many 2\,GHz observations.
The resulting spectral index is $\alpha = -1.9 \pm0.4$, where $S_{\nu}
\propto \nu^{\alpha}$.

\begin{figure}[t]
\begin{center}
\includegraphics[angle=0,scale=0.47]{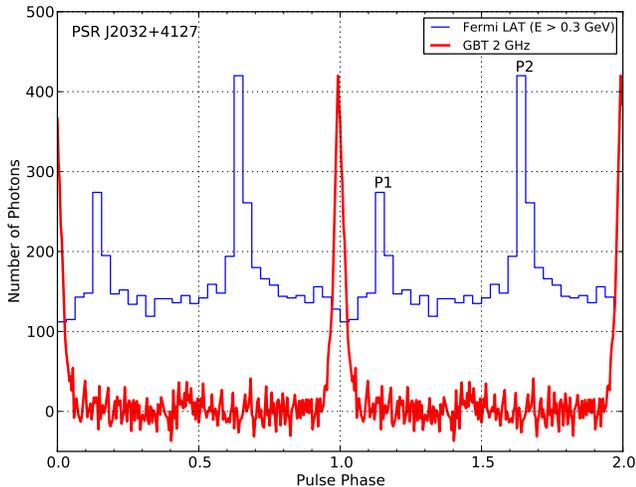}
\caption{\label{fig:psrbg}
Phase-aligned GBT and \fermi\ pulse profiles of \psrb.  The gamma-ray
peaks are modeled as Gaussians of, respectively, $\mbox{FWHM}/P = 0.026
\pm 0.003$ and $0.051 \pm 0.005$.  The radio profile is displayed with
an arbitrary intensity scale.  The radio and gamma-ray profiles are
displayed with, respectively, 256 and 32 bins per period.  Two full
rotations are shown.
}
\end{center} 
\end{figure}

\subsubsection{Gamma-ray and radio timing and position} \label{sec:timingb}

We are timing \psrb\ at the GBT.  The measured $\mbox{DM} =
(114.8\pm0.1)$\pccc, at $(l,b) = (80\fdg22,+1\fdg03)$, corresponds
to $d = 3.6$\,kpc, according to the \citet{cl02} model.  Hereafter,
we parameterize this distance by $d_{3.6} = d/(3.6\,\mbox{kpc})$.  The
average radio TOA uncertainty is 3 times smaller than the corresponding
gamma-ray value, but we have a radio data span of only 6 months compared
to 1 year for \fermi.  As a consequence, the best overall timing
solution is obtained from a joint fit.  Some rotational instability
is detectable in \psrb\ as timing noise.  This is parameterized
in TEMPO by the (non-stationary) second derivative of the rotation
frequency, $(-1.7\pm0.3)\times10^{-21}$\,s$^{-3}$.  The measured $\dot
P = 2.00\times10^{-14}$ implies $\tau_c = 0.11$\,Myr and $\dot E =
2.7\times10^{35}$\ergs.  The timing fit gives positional uncertainties
in R.A.\ and decl.\ of $0\farcs4$ and $0\farcs2$, respectively, although
timing noise may contribute a systematic error amounting to $\sim 1''$.

\begin{figure*}[t]
\centerline{
\hfill
\psfig{figure=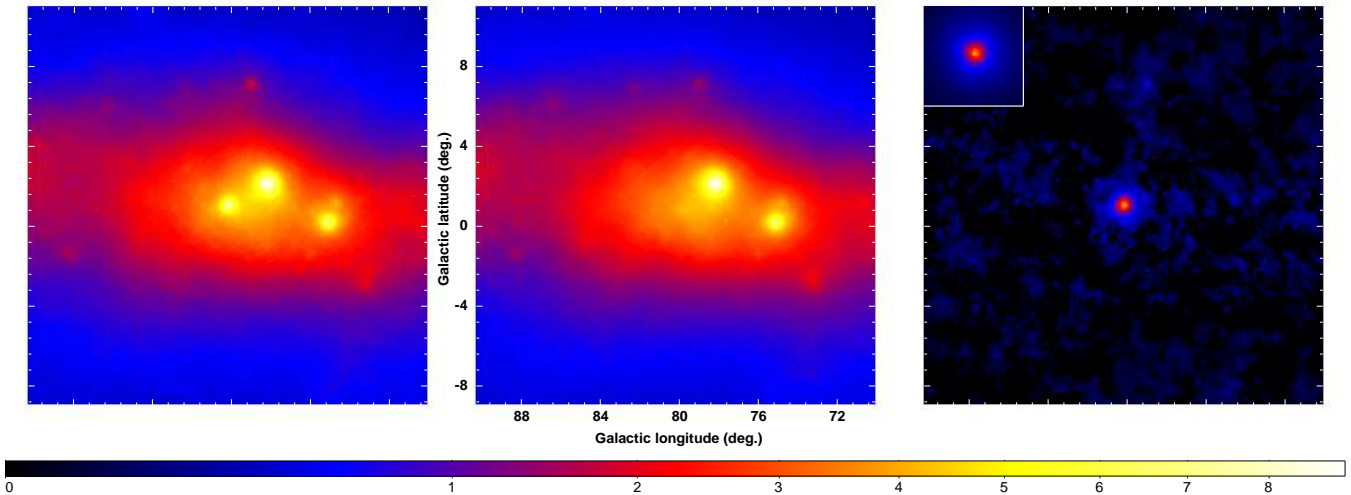,width=0.99\linewidth,angle=0}
\hfill
}
\caption{\label{fig:fermi}
The observed \fermi-LAT weighted counts map of the Cygnus region
($20\arcdeg\times20\arcdeg$ centered on \psrb).  The photons, selected
with the same time and zenith angle cuts used in the spectral analysis,
and with energies between 0.2 and 10\,GeV, were binned in energy and in
position with a data pixel size appropriate to the point spread function
(PSF) at the measured energy.  The counts in each pixel were divided
by the pixel area and the integration time and then interpolated onto a
$500\times500$ image grid using as weights the inverse angular separation
between image and data pixels out to a maximum separation equal to
about 80\% of the PSF.  This image cube was summed over energy, and the
color bar indicates the approximate rates in photons\,s$^{-1}$\,sr$^{-1}$
with a square-root scale.  {\em Left:} Photons selected by $0.08 < \phi <
0.20$ and $0.56 < \phi < 0.80$ for \psrb\ (pulsar ``on''). {\em Middle:}
Photons selected by $0.20 < \phi < 0.56$ and $0.80 < \phi < 1.08$ (pulsar
``off''), showing the background emission, including the two bright
pulsars J2021+4026 and J2021+3651.  {\em Right:} On--off difference map.
The only remaining source is \psrb.  The $5\arcdeg\times5\arcdeg$ inset
shows the PSF for a source with the same spectral energy dependence
as measured for \psrb, indicating that the pulsar is consistent with a
point source.
}
\end{figure*}

Radio imaging of this region was conducted by \citet{pmib07} and
\citet{mpib07}, including a 610\,MHz survey with the Giant Metrewave
Radio Telescope (GMRT), to identify potential counterparts for
the very-high-energy gamma-ray source \tev\ \citep{aab+02,aab+05}.
\citet{pmib07} noted that their GMRT source 5 coincides with star 213 in
the \citet[][hereafter MT91]{mt91} survey of massive stars in Cyg~OB2,
as well as with a \chandra\ X-ray source.  The latter associations are
explored further in \S~\ref{sec:xrayb}.  Here we conclude that \psrb\ and
GMRT source 5 are one and the same based on positional coincidence (see
Table~\ref{astrometry}), and because the $(0.5\pm0.1)$\,mJy flux of GMRT
source 5 at 610\,MHz is compatible within the uncertainties with the radio
spectrum of \psrb\ (\S~\ref{sec:radiob}).  Thus, the position of \psrb\
is known to $0\farcs5$ in each coordinate irrespective of timing noise.
We are able to limit possible systematic uncertainties in the astrometry
of the optical reference frame relative to the radio to $\approx 0\farcs1$
using the Tycho position of the star \cygf, which agrees with our optical
astrometry to this level.  Thus, the X-ray source coincides with the radio
source and the optical star to within $0\farcs6$, which is comparable
to their combined statistical and systematic uncertainties.

The timing fit also yields the phase offset between the radio and
gamma-ray profiles, which we show in Figure~\ref{fig:psrbg}.  The LAT
profiles are constructed from diffuse class photons selected from a region
of radius $0\fdg8$ around the best position over the time range 2008
August 4 to 2009 July 4.  We used only energies above 0.3\,GeV to help
reduce background contribution from nearby sources in this crowded field.
The phase alignment was done using {\em gtpphase} to assign phases to
each LAT photon according to the best timing model.  The phase spacing
between the two gamma-ray peaks ($\Delta = 0.50\pm0.01$) and the offset
between the radio and first (P1) LAT peak ($\delta = 0.15\pm0.01$) were
determined by fitting the unbinned photon phases to a two-Gaussian model.

We obtained the LAT position for \psrb\ with {\em ptlike} by selecting
photons above 0.125\,GeV from $0.12 < \phi < 0.20$ and $0.60 < \phi <
0.72$ (Figure~\ref{fig:psrbg}), and varying the position to maximize the
point source significance.  The resulting $\mbox{R.A.} = 20^{\rm h}32^{\rm
m}15\fs8$, $\mbox{decl.} = +41\arcdeg26'17''$, with a 95\%~C.L. error
radius of $1\farcm7$, is only $1\farcm2$ away from the pulsar timing
position.

\begin{deluxetable*}{lllll}[t]
\tablewidth{0.83\linewidth}
\tablecaption{Source Positions Coincident with Be Star \mt}
\tablehead{
\colhead{Source} & \colhead{Instrument} & \colhead{R.A. (J2000.0)} & \colhead{Decl. (J2000.0)} &
\colhead{Reference}
}
\startdata
\mt\      & MDM 2.4\,m RETROCAM         & 20 32 13.137(18)  & +41 27 24.28(20)  & This work \\
X-ray     & {\it Chandra} ACIS-I        & 20 32 13.143(24)  & +41 27 24.54(27)  & This work \\
Radio     & GMRT 610 MHz                & 20 32 13.092(39)  & +41 27 24.16(48)  & Mart\'i et al. (2007) \\
\psrb\    & {\it Fermi} LAT/GBT timing  & 20 32 13.07(4)    & +41 27 23.4(2)    & This work 
\enddata
\tablecomments{Units of right ascension are hours, minutes, and seconds,
and units of declination are degrees, arcminutes, and arcseconds.}
\label{astrometry}
\end{deluxetable*}

\subsubsection{Gamma-ray spectrum} \label{sec:gammab}

The Cygnus region contains several bright gamma-ray point sources
and strong diffuse emission that remains difficult to model (see
Figure~\ref{fig:fermi}).  However, as with \psra\ (\S~\ref{sec:gammaa}),
we found no evidence for point source emission off pulse, and we found
that we could measure the spectrum  of \psrb\ robustly by selecting only
photons from on-pulse phases (defined conservatively as being $0.08 <
\phi < 0.20$ and $0.56 < \phi < 0.80$).  We used the same time, energy,
and zenith angle cuts as for \psra\ (\S~\ref{sec:gammaa}), but with a
$20\arcdeg$ extraction radius, and modeled the bright nearby gamma-ray
pulsars J2021+3651 and J2021+4026 with an exponentially cut-off power-law
spectrum.  We find that an exponentially cut-off power law is preferred
over a simple power law with $9\,\sigma$ significance.  The best fit
parameters, $\Gamma = 1.1\pm0.2\pm0.2$ and $E_c = (3.0\pm0.6\pm0.7)$\,GeV,
give a photon flux above 0.1\,GeV of $(7\pm1\pm2)\times
10^{-8}$\,cm$^{-2}$\,s$^{-1}$ and $F_{\gamma} = (9\pm1\pm2)\times
10^{-11}$\,erg\,cm$^{-2}$\,s$^{-1}$.  For assumed isotropic emission,
$L_{\gamma} \approx 1.4\times10^{35}\,d_{3.6}^2\,\mbox{erg\,s$^{-1}$}
= 0.5\,d_{3.6}^2\,\dot E$.

As with \psra\ (\S~\ref{sec:gammaa}), the trailing peak has a harder
spectrum than P1: the P2/P1 ratio of baseline-subtracted maximum count
rate is 1.0 at 0.3--1\,GeV but rises to 1.5 above 1\,GeV, while the
ratio of total counts in the peaks changes from 2.3 to 2.8 in the same
energy ranges.  The highest energy photon likely detected so far from
this pulsar arrived at phase 0.642 (in P2) with energy 9.8\,GeV.

\subsubsection{X-ray and optical observations} \label{sec:xrayb}

\begin{figure*}[t]
\centerline{
\hfill
\psfig{figure=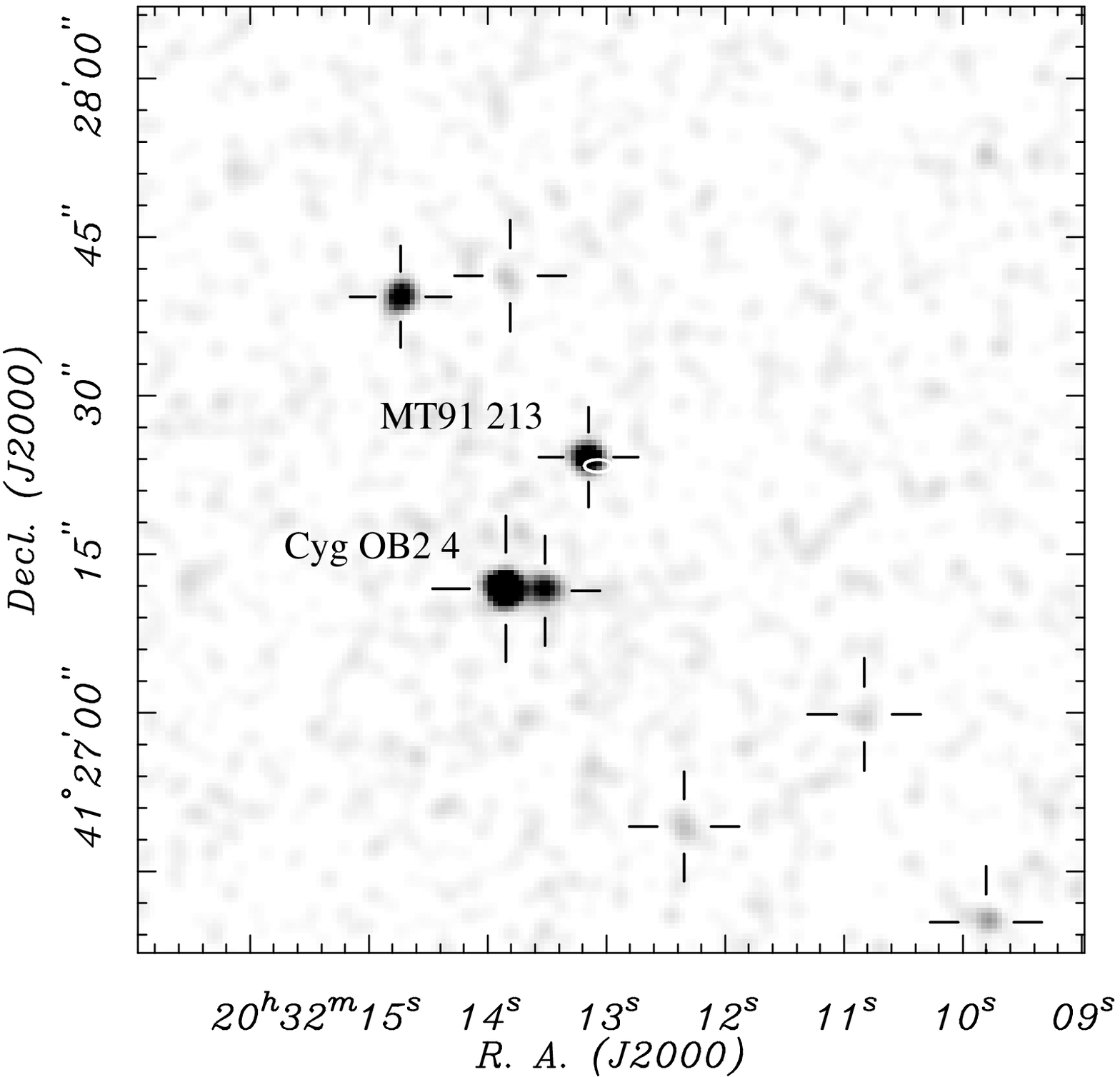,width=0.45\linewidth,angle=0}
\hfill
\psfig{figure=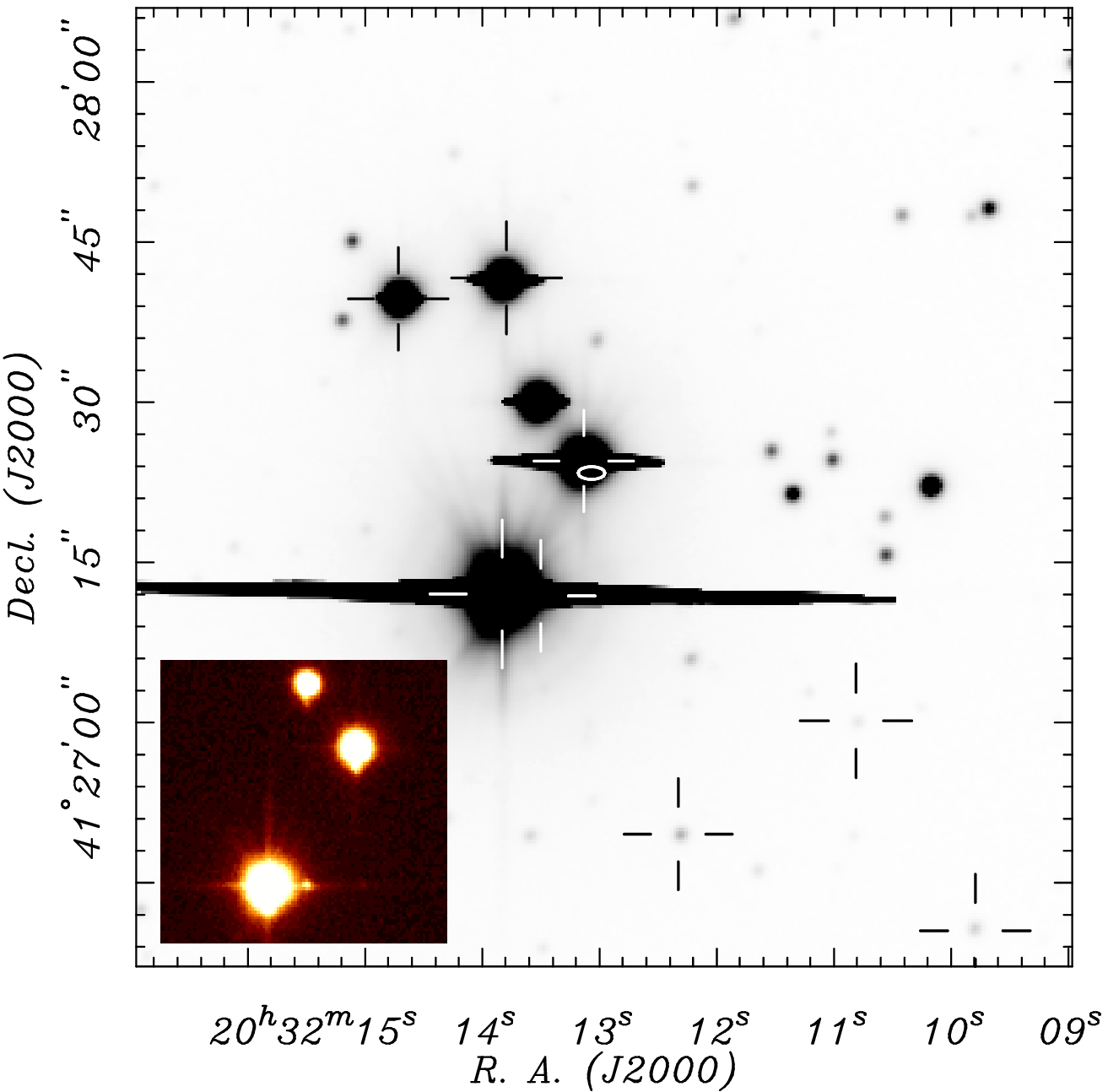,width=0.45\linewidth,angle=0}
\hfill
}
\caption{\label{fig:xrayopt}
{\em Left:}  Portion of \chandra\ image centered on the \psrb\
timing position, shown as a $3\,\sigma$ error ellipse.  The \chandra\
0.3--8\,keV ACIS-I image has been exposure-corrected and smoothed with
a $0\farcs5$ Gaussian kernel to highlight point sources.  All X-ray
point sources have an identified optical counterpart, shown on the
right panel, including \cxou, which is located $3\farcs7$ to the west
of the bright star \cygf.  The X-ray source coincident with the star
\mt\ (see Table~\ref{astrometry}) may be coming either from the star
or from \psrb.  {\em Right:}  An $R$-band image of the X-ray field,
taken with the MDM Observatory Hiltner 2.4\,m telescope.  The combined
exposure time is 20 minutes, and the seeing is $1\farcs0$ (see Mukherjee
et al.\ 2003\protect\nocite{mhg+03} for further details).  \cygf, which
masks the location of \cxou, \mt, and other bright stars are saturated.
Locations of \chandra\ sources, astrometrically corrected using this
image, are marked.  The inset shows another image that was specially
obtained in seeing of $0\farcs6$ (see \S~\ref{sec:xrayb}) to identify
the optical counterpart of \cxou, $3\farcs7$ to the west of \cygf.
}
\end{figure*}

\begin{figure}[t]
\begin{center}
\includegraphics[angle=270,scale=0.50]{f8.eps}
\caption{\label{fig:diffuse}
Diffuse X-ray emission in the vicinity of \psrb.  The point sources
have been removed from the \chandra\ 0.3--8\,keV ACIS-I image of
Figure~\ref{fig:xrayopt}, which is exposure-corrected and smoothed with
a Gaussian kernel of $\sigma = 14''$ to highlight diffuse emission.  This
image is a $7'\times7'$ zoom-in on Figure~3 from \citet{mgh07}.  The cross
is located at the pulsar timing position (see Figure~\ref{fig:xrayopt}).
The dot-dashed circle approximates the extent of the HEGRA source \tev.
}
\end{center} 
\end{figure}

\begin{figure*}
\hfil
\includegraphics[width=2.2in,angle=-90]{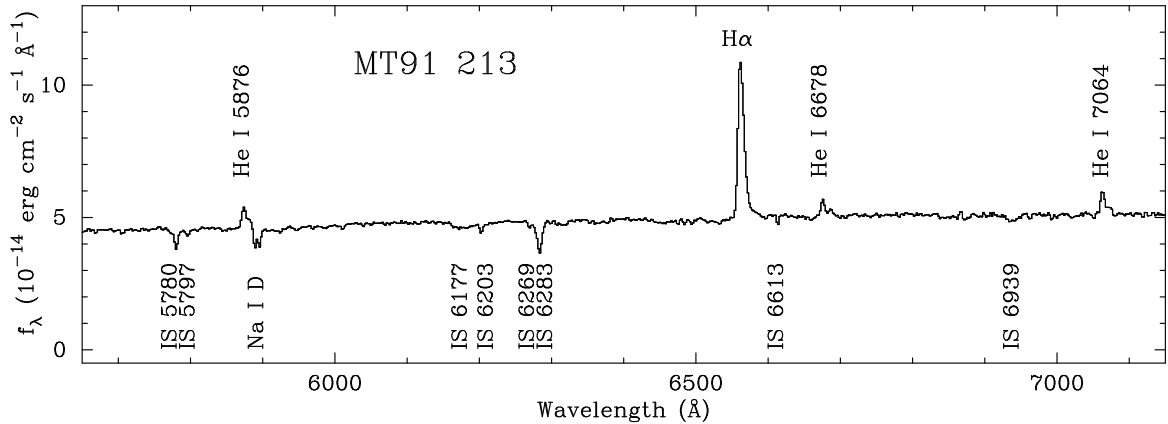}
\hfil
\caption{
Portion of the optical spectrum of \mt, showing emission lines that are
typical of Be stars.  The equivalent width of H$\alpha$ is $-12.6$\,\AA,
and the \ion{He}{1} lines have asymmetric peaks that are separated by
$\approx 500$\,km\,s$^{-1}$.  Interstellar absorption lines are marked
below the spectrum.
}
\label{fig:optspec}
\end{figure*}

\psrb\ is located in the direction of the massive \cyg\ stellar
association.  Several X-ray studies have been made of the region
containing the pulsar \citep{bdb+06,hhs+07,mhg+03,mgh07}, to identify
potential counterparts for \tev.  Here we have reanalyzed a 50\,ks
{\em Chandra X-ray Observatory} archival observation of the field
containing \psrb, obtained on 2004 July 12 with a time resolution of
3.2\,s \citep[observational details can be found in][]{bdb+06,mgh07}.
The pointing of this ACIS-I CCD observation placed \psrb\ $3'$ off-axis,
where the angular resolution is $\approx 1''$ FWHM.  We have examined
the area of the original radio gridding localization of \psrb\
(\S~\ref{sec:radiob}) and find several X-ray sources that are each
associated with an optical counterpart; the brightest is \cygf, an
$R=10.2$ O7~III((f)) star (see Figure~\ref{fig:xrayopt}).  Furthermore,
this is a region of enhanced diffuse hard X-ray emission identified
by \citet{mgh07}.  Figure~\ref{fig:diffuse} displays this $\sim 1'$
feature and shows the timing position of \psrb\ for reference.

This area was surveyed for optical counterparts using the MDM
Observatory Hiltner 2.4\,m telescope on 2002 August 23 \citep{mhg+03}.
We compared the X-ray sources and their optical counterparts in
Figure~\ref{fig:xrayopt} to register the X-ray image with a correction of
$-0\farcs16$ in R.A.\ and $+0\farcs31$ in decl.\ for the X-ray sources.
Thus, we confirm that an X-ray source coincides with the radio position
of \psrb/GMRT 5 and the star \mt\ to within $<0\farcs6$, as listed in
Table~\ref{astrometry}.

MT91 indicate a spectral type B0~Vp and $V=11.95$ for star 213.
We obtained a spectrum of \mt\ on 2009 June 13 using the MDM 2.4\,m
telescope with Modspec, covering the wavelength range 4220--7550\,\AA\
at 4\,\AA\ resolution.  Figure~\ref{fig:optspec} shows the red portion
of the spectrum, where emission lines of H$\alpha$ and \ion{He}{1}
are evident. The equivalent width of H$\alpha$ is $-12.6$\,\AA, and the
\ion{He}{1} lines have asymmetric peaks that are separated by $\approx
500$\,km\,s$^{-1}$, properties that are typical of Be stars.

Since their spectra covered only the blue region, where emission lines
are not obvious, MT91 did not report a Be classification for star 213.
Instead, they remarked that the absorption lines appear unusually
broad, and concluded that the star is multiple, thus the peculiar
classification.  Since all of the absorption lines in our spectrum are
filled in with emission, we have no additional evidence to support the
suggestion that \mt\ is a multiple system.  In addition, \citet{kkk+07}
did not find any evidence for binary motion among 10 spectra of \mt\
obtained between 1999 and 2004.  \citet{mt91} derive a color excess
$E(B-V) = 1.43$ for star 213, corresponding to $A_V=4.28$.  This optical
extinction would translate to X-ray column density $N_{\rm H} = 7.7 \times
10^{21}$\,cm$^{-2}$ using the conversion of \citet{pre95}.  We performed
spectral fits to the 83 X-ray photons detected from this source by
\chandra.  Power-law and Raymond-Smith plasma models fit equally well,
with photon index $\Gamma = 2.1 \pm 0.7$ and $kT = 4^{+9}_{-2}$\,keV,
respectively.  The fitted $N_{\rm H}$ is consistent with the optical
extinction to \mt.  The unabsorbed X-ray flux in the 0.5--10\,keV
band is $F_X \approx 3.2\times10^{-14}$\,erg\,cm$^{-2}$\,s$^{-1}$
for the power law. The corresponding luminosity is $L_X \approx
1.1\times10^{31}\,d_{1.7}^2$\,\mbox{erg\,s$^{-1}$}.

\citet{pmib07} suggested that the radio and X-ray sources coincident with
\mt\ are produced in a colliding wind binary.  While we now know that the
radio source is a pulsar, it is not clear what the relationship is between
the radio pulsar, the X-ray source, and the Be star.  The pulsar lacks any
evidence of binary motion in its timing.  The kinematic contribution to
the pulsar's frequency derivatives from its acceleration in orbit around
a $M_c \sim 20\,M_{\odot}$ companion (neglecting orbital eccentricity)
would be:
$$\left|{\dot f_k \over f}\right| < 1.2 \times 10^{-13}
\left(P_{\rm orb} \over 100\ {\rm yr}\right)^{-4/3}
\left(M_c \over 20\,M_{\odot}\right)^{1/3} \sin i\ \ \ {\rm s}^{-1},$$
$$\left|{\ddot f_k \over f}\right| < 3.2 \times 10^{-22}
\left(P_{\rm orb} \over 100\ {\rm yr}\right)^{-7/3}
\left(M_c \over 20\,M_{\odot}\right)^{1/3} \sin i\ \ \ {\rm s}^{-2}.$$
From Table~\ref{tab:parms}, we obtain $\dot f/f = -1.4 \times
10^{-13}$\,s$^{-1}$ and $\ddot f/f = -2.4 \times 10^{-22}$\,s$^{-2}$.
If the pulsar is a binary companion of the Be star, the orbital period
is probably $P_{\rm orb} \geq 100$\,yr in order not to exceed these
observed limits on the pulsar's acceleration.

Such long-period systems are not known to exist (the longest is 5
years), presumably because such a wide binary would not survive the
supernova kick.  Instead, the pulsar is probably superposed by chance on
the position of the star in this crowded field, in which case the X-ray
source may belong either to the radio pulsar or to the Be star, or both.
The degraded PSF at the off-axis location of the \chandra\ image does
not allow us to resolve this question.  But the X-ray luminosity of
the source is a fraction $\sim 4 \times 10^{-5}\,d_{1.7}^2$ of the
pulsar's $\dot E$, which is within the range of observed efficiencies
of young pulsars \citep[e.g.,][]{crg+06}.  It is also compatible with
the X-ray luminosities of single early Be stars, although on the high
end \citep{ber97,coh97}.

For completeness, we note that the location of one X-ray source
that lies $3\farcs7$ to the west of \cygf, \cxou, was saturated
in all of our optical images, as shown in the right main panel of
Figure~\ref{fig:xrayopt}.  Its corrected coordinates are $\mbox{R.A.} =
20^{\rm h}32^{\rm m}13\fs50$, $\mbox{decl.} = +41\arcdeg27'11\farcs8$,
with an error radius of $\approx 0\farcs3$.  We investigated the nature of
this source because it is the only other hard X-ray source (with half of
the 55 detected photons above 2\,keV) close to the pulsar timing position.

To identify \cxou\ it was necessary to obtain an unsaturated optical
image in good seeing.  Images of 1\,s exposure were acquired on the MDM
2.4\,m telescope on 2009 May 24 using RETROCAM \citep{mbd+05} and SDSS $r$
filter in seeing of $0\farcs6$.  The inset of Figure~\ref{fig:xrayopt}
(right) is the sum of 12 such exposures, which clearly reveals a star
to the west of  \cygf\ that we measure to fall within $0\farcs3$ of the
position of \cxou.  On 2009 May 25, we obtained an optical spectrum of
this star on the MDM 2.4\,m telescope, using a $1''$ wide slit that
cleanly isolated it.  The resulting spectrum is that of a Be star,
with an H$\alpha$ emission line as well as H and He absorption.

The optical spectra of \cxou\ and \cygf\ obtained on the same night
show interstellar absorption features that can be used to assess
their relative distances.  The equivalent widths of the features
agree to within $\sim 15\%$, comparable to their uncertainties, which
implies that \cxou\ is probably located in the Cyg~OB2 association.
There are too few X-ray counts from \cxou\ to obtain a spectrum, but
we can estimate its flux by assuming an absorbed power-law model.
\citet{mt91} derive a color excess $E(B-V) = 1.59$ for \cygf\
corresponding to $A_V=4.50$.  This optical extinction would translate
to X-ray column density $N_{\rm H} = 8 \times 10^{21}$~cm$^{-2}$
using the conversion of \citet{pre95}.  Assuming this $N_{\rm H}$ and
$\Gamma = 2.0$, the unabsorbed flux of \cxou\ in the 0.5--10\,keV band
is $F_X \approx 2.5\times10^{-14}$\,erg\,cm$^{-2}$\,s$^{-1}$.
The corresponding luminosity is $L_X \approx
9\times10^{30}\,d_{1.7}^2$\,\mbox{erg\,s$^{-1}$}.  X-ray luminosities
of Be stars decline sharply from early to late spectral type
\citep{ber97,coh97}.  The optical faintness of this star would seem to
indicate a late spectral type.  More precise spectral classification
of \cxou\ is needed to evaluate whether its high X-ray luminosity is
anomalous.  In any case, it is the second Be star discovered in the
vicinity of \psrb\ that is apparently not connected to it.  Curiously,
this star and \cygf\ are both closer to the centroid of the diffuse
emission in Figure~\ref{fig:diffuse} than is \psrb, even though that
may not have any physical significance.

\section{Discussion} \label{sec:disc} 

With the detection of radio pulsations from two of the 16 pulsars
recently discovered in blind searches of \fermi-LAT gamma-ray photons
\citep{abdo2}, we have taken a further step toward inferring the
fraction of radio-quiet gamma-ray-emitting pulsars.  Our non-detection
of radio pulsations from eight other LAT pulsars (\S~\ref{sec:obs})
is not constraining enough, owing to insufficient sensitivity:
minimum detectable flux densities were $S_{1.4} \sim 0.06$\,mJy for
the three observed by \citet{rhr+02} and $S_{1.4} \sim 0.2$\,mJy for
the remaining five, while very young radio pulsars have luminosities at
least as low as $L_{1.4} \equiv S_{1.4} d^2 \approx 0.5$\,mJy\,kpc$^2$
\citep{csl+02}, and the smallest measured pulsar luminosity is $L_{1.4}
\approx 0.03$\,mJy\,kpc$^2$ \citep[see][]{ymj99}.

\begin{deluxetable*}{lll}
\tablewidth{0.61\linewidth}
\tablecaption{\label{tab:parms} Measured and Derived Parameters for
PSRs~J1741--2054 and J2032+4127 }
\tablecolumns{3}
\tablehead{
\colhead{Parameter} &
\colhead{\psra }    &
\colhead{\psrb }
}
\startdata
Right ascension, R.A. (J2000.0)\tablenotemark{a}\dotfill &
  $17^{\rm h}41^{\rm m}57\fs27(8)$ & $20^{\rm h}32^{\rm m}13\fs07(4)$         \\
Declination, decl. (J2000.0)\tablenotemark{a}\dotfill    & 
  $-20\arcdeg53'13(34)''$          & $+41\arcdeg27'23\farcs4(2)$              \\
Galactic longitude, $l$ (deg.)\dotfill  &  6.42 & 80.22                       \\
Galactic latitude, $b$ (deg.)\dotfill   & +4.90 & +1.03                       \\
Spin period, $P$ (s)\dotfill & 0.41369996385(6) & 0.14324743146(2)            \\
Period derivative, $\dot P$ ($10^{-14}$)\dotfill & 1.694(3) & 2.0063(1)       \\
Epoch (MJD)\dotfill                     & 54826.0         & 54840.0           \\
Timing data span (MJD)\dotfill          & 54647--55004    & 54647--55033      \\
Timing residual, rms (ms)\dotfill       & 3.4             & 0.36              \\
Dispersion measure, DM (${\rm pc\,cm^{-3}}$)\dotfill & 4.7(1) & 114.8(1)      \\
Rotation measure, RM (rad\,m$^{-2}$)\dotfill & \nodata & $215\pm1$            \\
Flux density at 2\,GHz, $S_2$ (mJy)\dotfill  & \nodata \tablenotemark{b} & 
$0.12\pm0.03$                                                                 \\
Radio--gamma-ray profile offset, $\delta$ ($P$)\dotfill & 
  $0.29\pm0.02$                  & $0.15\pm0.01$                              \\
Gamma-ray profile peak-to-peak separation, $\Delta$ ($P$)\dotfill & 
  $0.23\pm0.02$\tablenotemark{c} & $0.50\pm0.01$                              \\
Gamma-ray ($>0.1$\,GeV) photon index, $\Gamma$\dotfill & 
  $1.4\pm0.1\pm0.1$              & $1.1\pm0.2\pm0.2$                          \\
Gamma-ray cut-off energy, $E_c$ (GeV)\dotfill & 
  $1.1\pm0.2\pm0.2$              & $3.0\pm0.6\pm0.7$                          \\
Photon flux ($>0.1$\,GeV) ($10^{-8}\,{\rm cm^{-2}\,s^{-1}}$) \dotfill & 
  $20\pm1\pm3$                   & $7\pm1\pm2$                                \\
Energy flux ($>0.1$\,GeV), $F_{\gamma}$ 
($10^{-11}\,{\rm erg\,cm^{-2}\,s^{-1}}$) \dotfill & 
  $12\pm1\pm2$                   & $9\pm1\pm2$                                \\
Spin-down luminosity, $\dot E$ (${\rm erg\,s^{-1}}$)\dotfill & 
  $9.4\times10^{33}$             & $2.7\times10^{35}$                         \\
Characteristic age, $\tau_c$ (Myr)\dotfill                        & 0.39& 0.11\\
Surface dipole magnetic field strength ($10^{12}$\,Gauss)\dotfill & 2.7 & 1.7 \\
Distance, $d$ (kpc)\tablenotemark{d}\dotfill                      & 0.4 & 3.6   
\enddata
\tablecomments{Values in parentheses are nominal TEMPO uncertainties on
the last digit for parameters determined in timing fits.  For gamma-ray
spectral parameters, the first uncertainty is statistical and the second
accounts for systematics. }
\tablenotetext{a}{The decl.\ of \psra\ is known with greater
precision than from this timing fit from a \swift\ observation
(see Figure~\ref{fig:swift}).  In the timing fit for \psrb\ we
account for rotational instability in the pulsar, parameterized
by the second derivative of spin frequency: $\ddot f =
-1.7(3)\times10^{-21}$\,s$^{-3}$. }
\tablenotetext{b}{The flux density received from \psra\ varies
greatly at 1.4\,GHz and 2\,GHz, due to interstellar scintillation (see
\S~\ref{sec:radioa}). }
\tablenotetext{c}{This is the separation between P1 and P3 (see
Figure~\ref{fig:psrag}). }
\tablenotetext{d}{Distances are estimated from the DM and the electron
density model of \citet{cl02}.  \psrb\ may be located at approximately
half this distance (see \S~\ref{sec:psrbd}). }
\end{deluxetable*}

The \fermi\ LAT bright source list \citep{abdo1} catalogs the 205 most
significant sources above 0.1\,GeV.  Due to the spatially varying Galactic
background, this is not a flux-limited sample, but we consider it here
as a crude proxy for an unbiased sample containing a substantial number
of pulsars.

A total of 30 known pulsars are represented in the list, including 15
discovered blindly in LAT data.  We exclude from consideration here
three (previously known) millisecond pulsars, since millisecond pulsars
could have different emission properties.  Of the remaining 27 pulsars,
13 are already known radio emitters, including the two reported here.
Four others can be considered to be radio quiet based on previous searches
\citep[see][]{bwa+04,hgc+04,hcg07,kl99}.  The status of the remaining
10 is still unclear, but given our success with PSRs~J1741--2054 and
J2032+4127, it is reasonable to suppose that some will eventually be
detected in the radio.  Therefore, among presently identified pulsars
in the LAT bright source list \citep{abdo1}, $\geq 50\%$ emit radio
pulsations.

In addition, however, nearly 50 of the 205 bright LAT sources are good
pulsar candidates (they are without known AGN associations and are not
variable).  Regardless of whether gamma-ray pulsations can be detected
from these, most of them are likely to be gamma-ray pulsars.  Thus, in
order to determine the true fraction of radio-quiet gamma-ray pulsars,
it is crucial to do sufficiently sensitive radio searches of most of
these 50 unidentified gamma-ray sources.

We now discuss PSRs~J1741--2054 and J2032+4127 in turn, for which
we have gathered many of the measured and derived parameters in
Table~\ref{tab:parms}.

\subsection{\psra } \label{sec:psrad}

The average radio flux density of \psra\ is uncertain due to
scintillation, but for our one detection at 1.4\,GHz, the nominal
luminosity is $L_{1.4} = 0.025\,d_{0.4}^2$\,mJy\,kpc$^2$, which is the
smallest of any detected radio pulsar.  Scintillation, which makes it
difficult to observe this pulsar, is a result of the very small DM and
thus small distance.  At $\sim 5$ times the distance (still at only
$\sim 2$\,kpc), a pulsar of such luminosity would not be detectable
with any existing telescope.  This is worth considering when discussing
``radio quiet'' neutron stars.

There are only eight pulsars known with DM smaller than that of \psra\
\citep[see][]{mhth05}\footnote{http://www.atnf.csiro.au/research/pulsar/psrcat}.
Two of them are exceedingly bright, and were discovered more than
40 years ago.  Four others are millisecond pulsars, for which such a
small DM is relatively easy to distinguish from zero in searches, and
thus from terrestrial radio interference.  Only the remaining two were
discovered in modern searches \citep{tnj+94,ymj99}.  Thus, while very
nearby pulsars are intrinsically rare, it is also the case that they
have proven difficult to detect in radio surveys, with implications for
the completeness of the known sample.

\psra\ was detected by EGRET as \egreta\ \citep{hbb+99}.  Its EGRET
flux above 0.1\,GeV is similar to the LAT flux (\S~\ref{sec:gammaa}).
With $P=413$\,ms, $\tau_c = 0.39$\,Myr, and $d \approx 0.4$\,kpc, \psra\
is reminiscent of the nearest middle-aged pulsars, Geminga ($P=237$\,ms;
$\tau_c = 0.34$\,Myr; $d = 0.25$\,kpc) and PSR~B0656+14 ($P=384$\,ms;
$\tau_c = 0.11$\,Myr; $d = 0.29$\,kpc), although it has only 1/4 of
their spin-down luminosity.

Among the known population of young (non-millisecond) pulsars, \psra\
has one of the smallest gamma-ray peak separations ($\Delta=0.23$) and
largest gamma-ray--radio lags ($\delta=0.29$).  Together, these values
are roughly in agreement with the expectation from ``outer-gap'' (OG)
gamma-ray beam models, which predict an inverse dependence of $\Delta$
on $\delta$ \cite[e.g.,][]{ry95}.  In OG models these parameters
depend principally on the viewing angle $\zeta$ (measured from the
rotation axis of the pulsar).  For \psra, the measured $(\Delta,\delta)$
indicate a smaller $\zeta$ than for most other known gamma-ray pulsars.
In detail, however, basic geometric models still cannot explain many of
the observed $(\Delta,\delta)$, including for \psra, let alone features
like the three peaks identified in this pulsar (Figure~\ref{fig:psrag}
and \S~\ref{sec:timinga}), and further ingredients are likely needed
\citep[see][]{wrwj09}.

In order to compute the flux of a gamma-ray pulsar averaged over the sky,
we need to know the geometry-dependent ``beaming'' factor $f_{\Omega}$
that corrects the phase-averaged flux measured by \fermi.  It is still
not possible to infer this precisely for a given pulsar, especially
when we do not possess independent geometric information (such as from
radio pulsar polarization measurements or from high-resolution X-ray
observations of pulsar wind nebulae).  Nevertheless, most promising
models --- which in addition to OG include ``two-pole caustic'' (TPC)
models \citep{dr03} --- predict large $f_{\Omega}$.  For the observed
profile characteristics of \psra, which resemble somewhat those of
PSR~B1706--44 \citep{ppp+09}, the OG and TPC models appear to suggest
that $f_{\Omega} \approx 0.5$--1 \citep{wrwj09}.

Based on the \fermi\ spectral parameters reported in
\S~\ref{sec:gammaa} for \psra, we obtain $L_{\gamma}
(>0.1\,{\rm GeV}) = 4\pi f_{\Omega} F_{\gamma} d^2 =
2.4\times10^{33}\,f_{\Omega}\,d_{0.4}^2\,\mbox{erg\,s$^{-1}$} =
0.25\,f_{\Omega}\,d_{0.4}^2\,\dot E$.  Plausibly $f_{\Omega} \approx
0.5$--1, which might eventually be improved with further modeling;
and $d_{0.4} \approx 1$.  \psra, which has the smallest $\dot E$ of
any established non-millisecond gamma-ray pulsar, has a high inferred
efficiency for converting spin-down luminosity into gamma-rays, $\eta
\equiv L_{\gamma} / \dot E = 0.25\,f_{\Omega}\,d_{0.4}^2$.   This is
in line with the trend that $\eta$ increases with decreasing $\dot E$
\citep{tbb+99}.  Constraining the distance with VLBI measurements will be
difficult, owing to the very small flux, but the proper motion could be
accessible via either VLBI or {\em HST} measurements.  Sensitive X-ray
observations will also prove very useful in constraining the properties
of this neutron star (see \S~\ref{sec:xraya}).

Non-millisecond gamma-ray pulsars typically have observed pulse
profiles with two distinct peaks.  In many such cases, the trailing
peak has a harder spectrum than P1.  Although \psra\ appears to have
three distinct peaks, it follows this trend in that its trailing peak
has a harder spectrum (see \S~\ref{sec:gammaa}).  One respect in which
\psra\ appears to stand out from non-millisecond pulsars is in having
the smallest spectral cut-off energy reported so far, $E_c = 1.1$\,GeV.
It remains to be determined whether this portends an incipient trend.

\subsection{\psrb } \label{sec:psrbd}

The radio luminosity of \psrb, scaled to the standard frequency of
1.4\,GHz using its estimated spectral index (\S~\ref{sec:radiob}),
is $L_{1.4} = 3\,d_{3.6}^2$\,mJy\,kpc$^2$.  While $\sim 100$ times
larger than that of \psra, this is still a small luminosity among
the observed young pulsar population \citep[see, e.g.,][]{cmg+02}.
The actual distance to the pulsar may differ significantly from the
3.6\,kpc estimated from the DM \citep{cl02}.  As we argue below, it may
be that $d_{3.6} \approx 0.5$.

As shown in Figure~\ref{fig:psrbr}, we have made polarimetric observations
of \psrb, and have measured Faraday rotation amounting to $\mbox{RM}
= +215$\,rad\,m$^{-2}$.  This implies an average Galactic magnetic
field along the line of sight, weighted by the free electron density,
of $2.3\,\mu$G, which is a typical Galactic value.  There are very few
pulsars with measured RM in the approximate direction of this pulsar
\citep{hml+06,abdo4}, so that this measurement cannot be put into
context and it does not provide a constraint on the pulsar distance.
Also, we have tried to obtain information on the pulsar geometry by
fitting a ``rotating vector model'' \citep{rc69a} to the swing with pulse
phase of the position angle of linear polarization (PA; top sub-panels
of Figure~\ref{fig:psrbr}).  Unfortunately the fits are unconstrained,
because of the limited longitude coverage and relatively shallow PA swing.
The radio profile of \psrb\ appears similar to those of several young
pulsars in being simple, relatively narrow, highly linearly polarized,
and in having little variation in PA \citep[see][]{jw06,wj08}.

The gamma-ray profile of \psrb\ (Figure~\ref{fig:psrbg}) is broadly
similar to those of several young pulsars, such as PSRs~J2021+3651
\citep{abdo4,hcg+08} and J0205+6449 \citep{abdo5}.  It has two narrow
gamma-ray peaks preceded by the radio pulse, with $(\Delta,\delta) =
(0.50,0.15)$.  These values can be approximately understood within the
context of both the OG and the TPC beam models.  In detail, however,
$\delta$ is somewhat larger than expected for the observed $\Delta$, given
the simplest geometric models \citep[see][]{wrwj09}.  This interpretive
problem arises as well for PSRs~J2021+3651 and J0205+6449.  Unlike for
\psra, these $(\Delta,\delta)$ indicate a large viewing angle $\zeta$
for \psrb, and a broad fan-like beam with correction factor $f_{\Omega}
\approx 1$.

Based on the spectrum of \psrb\ (\S~\ref{sec:gammab}),
we obtain $L_{\gamma} (>0.1\,{\rm GeV}) \approx
1.4\times10^{35}\,f_{\Omega}\,d_{3.6}^2\,{\rm erg\,s^{-1}}$.  With
$f_{\Omega} \approx 1$ as indicated above, $\eta \approx 0.5\,d_{3.6}^2$.
This is a very large nominal efficiency for a pulsar with $\dot E
= 2.7\times10^{35}$\ergs, which has an open field line voltage of
$10^{15}$\,V \citep[cf.][]{aro96b}.  Part of the answer may lie in a
smaller distance, and indeed it is plausible that $d_{3.6} \approx 0.5$
(see \S~\ref{sec:tev}).

\psrb\ is located in the Cygnus region, which contains several bright
gamma-ray point sources and strong spatially varying diffuse emission.
This leads to particular difficulty in modeling the spectrum of the
pulsar (see \S~\ref{sec:gammab} and Figure~\ref{fig:fermi}).  \psrb\
is marginally consistent with the location of the EGRET source \egretb\
\citep{hbb+99}, which is listed as \egr\ in the revised EGRET catalog of
\citet{cg08}.  We therefore consider \psrb\ as the likely identification
of the EGRET source.  The photon fluxes above 0.1\,GeV for \egretb\ and
\egr\ are, respectively, $73\pm7$ and $52\pm7$, in units of $10^{-8}\,{\rm
cm^{-2}\,s^{-1}}$.  In the \fermi\ bright source list \citep{abdo1},
\psrb\ corresponds to source 0FGL~J2032.2+4122, which has a flux
of $54\pm5$.  In \S~\ref{sec:gammab}, however, we report for \psrb\
a flux of $(7\pm1\pm2)\times 10^{-8}$\,cm$^{-2}$\,s$^{-1}$, or about
1/8 of that given for 0FGL~J2032.2+4122.  Compared to \citet{abdo1},
we have used significantly more data, an updated LAT response function,
an improved diffuse background model, considered only on-pulse emission
(see \S\S~\ref{sec:gammab} and \ref{sec:gammaa}), and modeled additional
point sources in the field.  Finally, the \citet{abdo1} spectral model
was a simple power law, generally inadequate to describe the spectral
breaks found in pulsars.  Together, these differences presumably account
for the large discrepancy in photon flux.  The new, much smaller,
flux value nevertheless implies an unreasonably large efficiency $\eta
\approx 0.5\,d_{3.6}^2$.  We believe that for \psrb\ our flux estimate
is the most reliable one available at this point, but still subject
to improvement as the modeling of the complex Cygnus region evolves.
And we also suggest that a reasonable conversion efficiency for \psrb\
is most naturally achieved with $d_{3.6} \approx 0.5$.

The trailing peak of \psrb\ has a harder spectrum than P1
(\S~\ref{sec:gammab}).  This behavior is similar to that observed in
PSRs~J2021+3651 and J0205+6449 \citep{abdo4,abdo5}, which as already
noted have pulse profiles broadly similar to \psrb, but have $\dot E$
that are larger by factors of 10 and 100, respectively.  This gathering
trend now applies to many known gamma-ray pulsars with distinct peaks,
from \psra\ (\S~\ref{sec:gammaa}) to the Crab \citep{tho01b}, spanning
a factor of 5000 in $\dot E$.

\subsubsection{The (formerly) unidentified \tev } \label{sec:tev}

\tev\ was discovered serendipitously by the HEGRA system of Atmospheric
Cherenkov Telescopes in observations of the Cygnus region over the
period 1999--2001 \citep{aab+02}.  It was the first unclassified TeV
source, and its origin remained undetermined until now.  Analysis of
combined HEGRA data from 1999--2002 gave a final position centroid for
the extended TeV source of $\mbox{R.A.} = 20^{\rm h}31^{\rm m}57\fs0 \pm
6\fs2_{\rm stat} \pm 13\fs7_{\rm sys}$, $\mbox{decl.} = +41\arcdeg29'57''
\pm 1\farcm1_{\rm stat} \pm 1\farcm0_{\rm sys}$, and a Gaussian radius
of $\sigma = 6\farcm2 \pm 1\farcm2_{\rm stat} \pm 0\farcm9_{\rm sys}$
\citep{aab+05}.  The HEGRA-measured photon flux above 1\,TeV is $(6.9
\pm 1.8) \times 10^{-13}$\,cm$^{-2}$\,s$^{-1}$ with $\Gamma = 1.9
\pm 0.3$.  MAGIC detected \tev\ at a best position of $\mbox{R.A.} =
20^{\rm h}32^{\rm m}20^{\rm s} \pm 11^{\rm s}_{\rm stat} \pm 11^{\rm
s}_{\rm sys}$, $\mbox{decl.} = +41\arcdeg30'36'' \pm 1\farcm2_{\rm stat}
\pm 1\farcm8_{\rm sys}$, and radius $\sigma = 6\farcm0 \pm 1\farcm7_{\rm
stat} \pm 0\farcm6_{\rm sys}$ \citep{aaa+08}.  The flux measured by
MAGIC above 1\,TeV is $4.5 \times 10^{-13}$\,cm$^{-2}$\,s$^{-1}$ with
photon index $\Gamma = 2.0 \pm 0.3$.  The Whipple Observatory detected
a source of luminosity $L_{\gamma} = 4 \times 10^{33}\,d_{1.7}^2$\ergs,
assuming a Crab-like spectrum \citep{kab+07}, which is about a factor
of 2 greater than measured by HEGRA, but probably consistent given the
uncertainty in the spectrum.  The Whipple position and extent of the
source are marginally consistent with those of HEGRA.  Milagro detects a
diffuse source in a $3\arcdeg \times 3\arcdeg$ region centered on \tev,
that exceeds the HEGRA flux by a factor of 3 \citep{aab+07}.  This region
may contain multiple sources; the Milagro excess at the location of
the pulsar is $7.6\,\sigma$ \citep{abdo7} and is consistent with an
extrapolation of the HEGRA spectrum.  None of the experiments have found
strong evidence for flux variability of \tev\ in the period 1999--2005.

Pulsar wind nebulae (PWNe) comprise the largest class of Galactic
TeV sources\footnote{http://www.mppmu.mpg.de/$\sim$rwagner/sources},
and all but the youngest, with ages of $\sim 10^3$\,yr, are spatially
extended.  \psrb\ is located within the $1\,\sigma$ extent of \tev\
(see Figure~\ref{fig:diffuse}), only $4'$ from its HEGRA centroid.
We therefore propose its PWN as the source of \tev.  Further support
for the association of \psrb\ with \tev\ comes from comparing their
properties with those of other TeV PWNe.

First, however, we discuss the implications of the apparent coincidence of
\psrb\ with the massive young star cluster \cyg.  Distance estimates to
\cyg\ range over 1.45--1.7\,kpc \citep{han03,mt91}, while the nominal DM
distance of \psrb\ is 3.6\,kpc.  Considering the possibility of unmodeled
local enhancements in the electron density that may cause the DM model to
overestimate the pulsar distance, we do not rule out that \psrb\ could
be colocated with \cyg\ (which, as noted in \S~\ref{sec:psrbd}, would
imply a far more reasonable pulsar GeV conversion efficiency $\eta$).
\psrb\ is projected $15'$ from the center of the cluster, which has a
half-light radius of $13'$ and a diameter of $\sim 2\arcdeg$ as determined
from 2MASS star counts \citep{kno00}.

Conventional wisdom holds that no neutron stars have been born in
\cyg\ because the age of the majority of its stars, determined from
isochrone fitting, is only 2--2.5\,Myr \citep{han03,nmhc08}.  At this
age, stars of $M < 35\ M_{\odot}$ are still close to the main sequence,
and no supernovae have occurred.  However, there is evidence for earlier
episodes of star formation in the immediate neighborhood, such as the
$\sim 5$--7\,Myr-old A stars discovered by \citet{dgis08} within a
$1\arcdeg$ radius of \cyg\ and at the same distance.  Older OB stars
coincident with \cyg\ are also discussed by \citet{han03}.  Therefore,
we cannot rule out that the progenitor star of \psrb\ was born in a recent
episode of star formation at the same distance as \cyg, while the typical
neutron star velocity of $\sim 250$\,km\,s$^{-1}$ would allow it to have
travelled $\sim 1\arcdeg$ at that distance in $10^5$\,yr.

The offset of \tev\ from the center of \cyg\ is most easily explained by
the existence and location of \psrb, i.e., that \psrb\ is the source of
\tev, and not, e.g., winds from O stars in the cluster.  In the latter
case, it would be difficult to understand why the TeV emission is not
more widely distributed among the dozens of O stars of \cyg, and centered
on the cluster, which is quite spherical in 2MASS \citep{kno00}. This
argument for associating \psrb\ with \tev\ applies whether or not \psrb\
is actually in the cluster.

In order to compare the efficiency of TeV gamma-ray production with
those of other pulsars listed by \citet{gcd+08}, we integrate the
HEGRA spectrum over 0.3--30\,TeV, giving $F(0.3-30\ {\rm TeV}) = 5.1
\times 10^{-12}$\,erg\,cm$^{-2}$\,s$^{-1}$, $L_{\gamma} = 1.8 \times
10^{33}\,d_{1.7}^2$\ergs, and $\epsilon \equiv L_{\gamma}/\dot E =
0.007\,d_{1.7}^2$.  This is consistent with the range of efficiencies
$\epsilon = 10^{-4}$--0.11 found for other PWNe and candidates by
\citet{gcd+08} \citep[see also][]{hng+08}.  Even at $d = 3.6$\,kpc,
$\epsilon = 0.03$ is not exceptional.  Also typical for a PWN is
the flux of the diffuse X-rays within $1'$ of \psrb, $\sim 1 \times
10^{-13}$\,erg\,cm$^{-2}$\,s$^{-1}$ in the 0.5--10\,keV band \citep[see
Figure~\ref{fig:diffuse};][]{mgh07}, which corresponds to $L_X/\dot E
\sim 1 \times 10^{-4}\,d_{1.7}^2$.  Therefore, we approximate the ratio
$L_{\gamma}/L_X \sim 50$, which is comparable to the ratio for other
older pulsars \citep{mfg+09}.  However, we do not see any evidence for
a larger X-ray nebula covering almost the entire HEGRA source that was
claimed by \citet{hhs+07} using \xmm\ images, with an order of magnitude
higher flux than our small nebula.

The spin-down luminosity of \psrb\ stands out as the smallest of
any pulsar identified with a TeV source; the others have $\dot E >
10^{36}$\ergs.  One possible exception is PSR~J1702--4128 with $\dot E
= 3.4 \times 10^{35}$\ergs, if it is associated with HESS~J1702--420.
The latter identification is problematic, however \citep{aab+08},
because it would require $\epsilon \sim 0.7$ at the estimated distance
of 5\,kpc, and a rather extreme asymmetry for the TeV emission, peaking
more than $0\fdg6$ from the pulsar.  Nevertheless, the existence of
still unidentified TeV sources allows the possibility that new pulsar
identifications may be found that are less energetic than the already
known pulsar counterparts, and \tev\ is probably one such example.

Shortly after the discovery of \tev, \citet{bed03} proposed that
\egretb\ is a Vela-type pulsar in \cyg, and that \tev\ is its PWN.
Using a time-dependent model for the PWN that includes both hadronic and
leptonic processes, he assumed a birth period of 2\,ms, $B \sim 6 \times
10^{12}$\,G, and present pulsar parameters $P = 210$\,ms and $\tau_c =
2 \times 10^4$\,yr, corresponding to $\dot E = 2.3 \times 10^{35}$\ergs.
The latter is close to the measured value for \psrb\ of $\dot E = 2.7
\times 10^{35}$\ergs, although the measured characteristic age of $1.1
\times 10^5$\,yr suggests that the time-dependence of the model is not
well constrained.  For example, \psrb\ may be too old for its initial
spin energy to have had much influence on its present TeV luminosity.
See \citet{dd08} for a discussion of relevant timescales.

\citet{bcd+08} point to a possible shell-like arrangement of predominantly
non-thermal radio emission, centered on the TeV source, in VLA images
\citep[see also][]{pmib07}.  While it is not even clear that this is
a single coherent structure, it is unlikely to be the remnant of the
supernova that gave birth to the $\sim 10^5$\,yr-old \psrb, as its radius
of 3\,pc at the distance of \cyg\ would be too small.  Using higher
resolution VLA data, \citet{mpib07} do not confirm a supernova remnant
nature for this emission.

Although \psrb\ is consistent in position with the Be star \mt, there
is no other evidence that they are a binary pair such as the prototype
gamma-ray emitting binary PSR~B1259--63, which has an orbital period
of 3.4\,yr.  Nevertheless, if the proximity of \psrb\ to \cyg\ is real
rather than apparent, it is likely that the local radiation background
from the massive star cluster enhances inverse Compton TeV emission from
the PWN, which might otherwise not have been detectable.  Although the
total luminosity from massive stars in \cyg\ is uncertain, estimates
of $\sim 100$ O stars lead to a luminosity of $\sim 10^{41}$\ergs,
and a photon energy density of $\sim 10^2$\,eV\,cm$^{-3}$, which at the
location of \psrb\ may manifest itself mainly as reprocessed IR emission
from dust (compared with 0.26\,eV\,cm$^{-3}$ for the cosmic microwave
background).  Detailed modeling, including inverse Compton scattering
in the Klein-Nishina regime, will be needed to evaluate the possible
TeV emission using this enhanced background radiation, and could help
constrain the distance between \psrb\ and \cyg.

\section{Conclusions} \label{sec:concl}

We report radio pulsar detections of two \fermi\ sources previously known
only as gamma-ray pulsars, as well as details of their gamma-ray spectra,
and probable identifications of their X-ray counterparts.  Gamma-ray
efficiencies are estimated using their radio dispersion-measure distances.
Both pulsars have hard power-law gamma-ray spectra with exponential
cut-offs in the GeV band, and large efficiencies relative to spin-down
power of $\sim 0.2$, assuming isotropic emission.  The phase offsets
between their radio and gamma-ray pulses follow trends observed in other
pulsars, probably indicating their viewing geometry, and consistent
with outer magnetosphere models, although more physics must be added
to existing calculations to model the detailed structure and spectral
evolution of the observed light curves.

\psra\ has a low $\dot E = 9.4 \times 10^{33}$\ergs, and its
characteristic age of 0.4\,Myr is compatible with the soft X-ray
spectral characteristics of its putative counterpart, inferred to be
surface thermal emission plus a non-thermal component.  With a radio
luminosity smaller than that of any other known radio pulsar, \psra\ at
a distance of 0.4\,kpc is approximately twice as far as the radio-quiet,
middle-aged gamma-ray pulsar Geminga, which it resembles in its spin-down
parameters and X-ray properties.  \psra\ begins to answer the question
of where are the other Gemingas.

\psrb\ is a more energetic pulsar, with $\dot E = 2.7\times10^{35}$\ergs,
and is brighter in radio.  A precise timing position derived from a joint
fit to \fermi\ and GBT data confines the pulsar position to a region of
diffuse X-ray emission previously identified in a \chandra\ image, which
is now presumed to be its PWN.  The location of \psrb\ within the extent
of the diffuse HEGRA source \tev\ solves the 10-year-old mystery of the
nature of this, the first unidentified TeV source.  \psrb\ is probably one
of the least energetic pulsars powering TeV PWNe, which are now known to
be the most numerous type of Galactic TeV source.  The location of \psrb\
projected close to the core of the massive, young stellar association
\cyg\ at a distance of 1.5--1.7\,kpc suggests that this is its true
distance, rather than 3.6\,kpc estimated from the radio pulsar dispersion.

\acknowledgements

We thank profusely all the \fermi\ team members who have built and who
operate the magnificent spacecraft and LAT experiment.  The GBT is
operated by the National Radio Astronomy Observatory, a facility of
the National Science Foundation operated under cooperative agreement
by Associated Universities, Inc.  The Parkes Observatory is part of the
Australia Telescope, which is funded by the Commonwealth of Australia for
operation as a National Facility managed by CSIRO.  We are grateful to
the \swift\ project scientist and staff for the observation of the LAT
\psra\ field, and to John Thorstensen for taking the MDM optical spectrum.

The \fermi\ LAT Collaboration acknowledges generous ongoing support
from a number of agencies and institutes that have supported both the
development and the operation of the LAT as well as scientific data
analysis.  These include the National Aeronautics and Space Administration
and the Department of Energy in the United States, the Commissariat \`a
l'Energie Atomique and the Centre National de la Recherche Scientifique /
Institut National de Physique Nucl\'eaire et de Physique des Particules in
France, the Agenzia Spaziale Italiana and the Istituto Nazionale di Fisica
Nucleare in Italy, the Ministry of Education, Culture, Sports, Science
and Technology (MEXT), High Energy Accelerator Research Organization
(KEK) and Japan Aerospace Exploration Agency (JAXA) in Japan, and the
K.~A.~Wallenberg Foundation, the Swedish Research Council and the Swedish
National Space Board in Sweden.

{\em Facilities:}  \facility{CXO (ACIS-I)}, \facility{Fermi (LAT)},
\facility{GBT (BCPM, GUPPI)}, \facility{Hiltner (RETROCAM)},
\facility{Parkes (PMDAQ)}, \facility{Swift (XRT)}


\begin{thebibliography}{70}
\expandafter\ifx\csname natexlab\endcsname\relax\def\natexlab#1{#1}\fi

\bibitem[{{Abdo} {et~al.}(2007){Abdo}, {Abdo}, \& {Abdo}}]{aab+07}
{Abdo}, A.~A., et~al. 2007, \apj, 658, L33

\bibitem[{{Abdo} {et~al.}(2009{\natexlab{a}}){Abdo}, {Abdo}, \& {Abdo}}]{abdo4}
---. 2009{\natexlab{a}}, ApJ, 700, 1059

\bibitem[{{Abdo} {et~al.}(2009{\natexlab{b}}){Abdo}, {Abdo}, \& {Abdo}}]{abdo5}
---. 2009{\natexlab{b}}, ApJ, 699, L102

\bibitem[{{Abdo} {et~al.}(2009{\natexlab{c}}){Abdo}, {Abdo}, \& {Abdo}}]{abdo1}
---. 2009{\natexlab{c}}, ApJS, 183, 46

\bibitem[{{Abdo} {et~al.}(2009{\natexlab{d}}){Abdo}, {Abdo}, \& {Abdo}}]{abdo3}
---. 2009{\natexlab{d}}, ApJ, 696, 1084

\bibitem[{{Abdo} {et~al.}(2009{\natexlab{e}}){Abdo}, {Abdo}, \& {Abdo}}]{abdo7}
---. 2009{\natexlab{e}}, ApJ, 700, L127

\bibitem[{{Abdo} {et~al.}(2009{\natexlab{f}}){Abdo}, {Abdo}, \& {Abdo}}]{abdo2}
---. 2009{\natexlab{f}}, Science, 325, 840

\bibitem[{{Abdo} {et~al.}(2009{\natexlab{g}}){Abdo}, {Abdo}, \& {Abdo}}]{abdo10}
---. 2009{\natexlab{g}}, Science, 325, 848

\bibitem[{{Aharonian} {et~al.}(2002){Aharonian}, {Aharonian}, \&
  {Aharonian}}]{aab+02}
{Aharonian}, F., et~al. 2002, A\&A, 393, L37

\bibitem[{{Aharonian} {et~al.}(2005){Aharonian}, {Aharonian}, \&
  {Aharonian}}]{aab+05}
---. 2005, A\&A, 431, 197

\bibitem[{{Aharonian} {et~al.}(2008){Aharonian}, {Aharonian}, \&
  {Aharonian}}]{aab+08}
---. 2008, A\&A, 477, 353

\bibitem[{{Albert} {et~al.}(2008){Albert}, {Albert}, \& {Albert}}]{aaa+08}
{Albert}, J., et~al. 2008, ApJ, 675, L25

\bibitem[{{Arons}(1996)}]{aro96b}
{Arons}, J. 1996, \aaps, 120, 49

\bibitem[{Atwood {et~al.}(2009)Atwood, Atwood, \& Atwood}]{atwood}
Atwood, W.~B., et~al. 2009, ApJ, 697, 1071

\bibitem[{Backer {et~al.}(1997)Backer, Dexter, Zepka, D., Wertheimer, Ray, \&
  Foster}]{bdz+97}
Backer, D.~C., Dexter, M.~R., Zepka, A., D., N., Wertheimer, D.~J., Ray, P.~S.,
  \& Foster, R.~S. 1997, PASP, 109, 61

\bibitem[{{Becker} {et~al.}(2004){Becker}, {Weisskopf}, {Arzoumanian},
  {Lorimer}, {Camilo}, {Elsner}, {Kanbach}, {Reimer}, {Swartz}, {Tennant}, \&
  {O'Dell}}]{bwa+04}
{Becker}, W., et~al. 2004, \apj, 615, 897

\bibitem[{{Bednarek}(2003)}]{bed03}
{Bednarek}, W. 2003, \mnras, 345, 847

\bibitem[Bergh\"ofer et al.(1997)]{ber97}
Bergh\"ofer, T. W., Schmitt, J. H. M. M., Danner, R., \& Cassinelli, J. P.
1997, \aap, 322, 167

\bibitem[{{Butt} {et~al.}(2008){Butt}, {Combi}, {Drake}, {Finley}, {Konopelko},
  {Lister}, {Rodriguez}, \& {Shepherd}}]{bcd+08}
{Butt}, Y.~M., {Combi}, J.~A., {Drake}, J., {Finley}, J.~P., {Konopelko}, A.,
  {Lister}, M., {Rodriguez}, J., \& {Shepherd}, D. 2008, \mnras, 385, 1764

\bibitem[{{Butt} {et~al.}(2006){Butt}, {Drake}, {Benaglia}, {Combi}, {Dame},
  {Miniati}, \& {Romero}}]{bdb+06}
{Butt}, Y.~M., {Drake}, J., {Benaglia}, P., {Combi}, J.~A., {Dame}, T.,
  {Miniati}, F., \& {Romero}, G.~E. 2006, \apj, 643, 238

\bibitem[{Camilo {et~al.}(2002{\natexlab{a}})Camilo, Manchester, Gaensler,
  Lorimer, \& Sarkissian}]{cmg+02}
Camilo, F., Manchester, R.~N., Gaensler, B.~M., Lorimer, D.~L., \& Sarkissian,
  J. 2002{\natexlab{a}}, ApJ, 567, L71

\bibitem[{{Camilo} {et~al.}(2006){Camilo}, {Ransom}, {Gaensler}, {Slane},
  {Lorimer}, {Reynolds}, {Manchester}, \& {Murray}}]{crg+06}
{Camilo}, F., {Ransom}, S.~M., {Gaensler}, B.~M., {Slane}, P.~O., {Lorimer},
  D.~R., {Reynolds}, J., {Manchester}, R.~N., \& {Murray}, S.~S. 2006, ApJ,
  637, 456

\bibitem[{Camilo {et~al.}(2002{\natexlab{b}})Camilo, {Stairs}, {Lorimer},
  {Backer}, {Ransom}, {Klein}, {Wielebinski}, {Kramer}, {McLaughlin},
  {Arzoumanian}, \& {M{\" u}ller}}]{csl+02}
Camilo, F., et~al. 2002{\natexlab{b}}, ApJ, 571, L41

\bibitem[{{Casandjian} \& {Grenier}(2008)}]{cg08}
{Casandjian}, J.-M., \& {Grenier}, I.~A. 2008, A\&A, 489, 849

\bibitem[Cohen et al.(1997)]{coh97}
Cohen, D. H., Cassinelli, J. P., \& MacFarlane, J. J. 1997, \apj, 487, 867

\bibitem[{{Cordes} \& {Lazio}(2002)}]{cl02}
{Cordes}, J.~M., \& {Lazio}, T.~J.~W. 2002, arXiv:astro-ph/0207156

\bibitem[{{de Jager} \& {Djannati-Ata{\"i}}(2008)}]{dd08}
{de Jager}, O.~C., \& {Djannati-Ata{\"i}}, A. 2008, in Neutron Stars and
  Pulsars: 40 Years After Their Discovery, ed. W.~Becker (Berlin: Springer),
  451

\bibitem[{{Drew} {et~al.}(2008){Drew}, {Greimel}, {Irwin}, \& {Sale}}]{dgis08}
{Drew}, J.~E., {Greimel}, R., {Irwin}, M.~J., \& {Sale}, S.~E. 2008, \mnras,
  386, 1761

\bibitem[{{Dyks} \& {Rudak}(2003)}]{dr03}
{Dyks}, J., \& {Rudak}, B. 2003, ApJ, 598, 1201

\bibitem[{{Gallant} {et~al.}(2008){Gallant}, {Carrigan}, {Djannati-Ata{\"i}},
  {Funk}, {Hinton}, {Hoppe}, {de Jager}, {Kh{\'e}lifi}, {Komin}, {Kosack},
  {Lemi{\`e}re}, \& {Masterson}}]{gcd+08}
{Gallant}, Y.~A., et~al. 2008, in AIP Conf.
  Ser. 983, 40 Years of Pulsars: Millisecond Pulsars, Magnetars and More,
  ed. C.~{Bassa}, Z.~{Wang}, A.~{Cumming}, \& V.~M. {Kaspi} 
  (Berlin: Springer), 195

\bibitem[{{Halpern} {et~al.}(2008){Halpern}, {Camilo}, {Giuliani}, {Gotthelf},
  {McLaughlin}, {Mukherjee}, {Pellizzoni}, {Ransom}, {Roberts}, \&
  {Tavani}}]{hcg+08}
{Halpern}, J.~P., et~al. 2008, \apj, 688, L33

\bibitem[{{Halpern} {et~al.}(2007){Halpern}, {Camilo}, \& {Gotthelf}}]{hcg07}
{Halpern}, J.~P., {Camilo}, F., \& {Gotthelf}, E.~V. 2007, \apj, 668, 1154

\bibitem[{{Halpern} {et~al.}(2004){Halpern}, {Gotthelf}, {Camilo}, {Helfand},
  \& {Ransom}}]{hgc+04}
{Halpern}, J.~P., {Gotthelf}, E.~V., {Camilo}, F., {Helfand}, D.~J., \&
  {Ransom}, S.~M. 2004, \apj, 612, 398

\bibitem[{Han {et~al.}(2006)Han, Manchester, Lyne, Qiao, \& van
  Straten}]{hml+06}
Han, J.~L., Manchester, R.~N., Lyne, A.~G., Qiao, G.~J., \& van Straten, W.
  2006, ApJ, 642, 868

\bibitem[{{Hanson}(2003)}]{han03}
{Hanson}, M.~M. 2003, \apj, 597, 957

\bibitem[{{Harding} {et~al.}(2007){Harding}, {Grenier}, \& {Gonthier}}]{hgg07}
{Harding}, A.~K., {Grenier}, I.~A., \& {Gonthier}, P.~L. 2007, Ap\&SS, 309, 221

\bibitem[{Hartman {et~al.}(1999)Hartman, Bertsch, Bloom, Chen, Deines-Jones,
  Esposito, Fichtel, Friedlander, Hunter, McDonald, Sreekumar, Thompson, Jones,
  Lin, Michelson, Nolan, Tompkins, Kanbach, Mayer-Hasselwander, M\"ucke, Pohl,
  Reimer, Kniffen, Schneid, von Montigny, Mukherjee, \& Dingus}]{hbb+99}
Hartman, R.~C., et~al. 1999, ApJS, 123, 79

\bibitem[{{Hessels} {et~al.}(2008){Hessels}, {Nice}, {Gaensler}, {Kaspi},
  {Lorimer}, {Champion}, {Lyne}, {Kramer}, {Cordes}, {Freire}, {Camilo},
  {Ransom}, {Deneva}, {Bhat}, {Cognard}, {Crawford}, {Jenet}, {Kasian},
  {Lazarus}, {van Leeuwen}, {McLaughlin}, {Stairs}, {Stappers}, \&
  {Venkataraman}}]{hng+08}
{Hessels}, J.~W.~T., et~al. 2008, ApJ, 682, L41

\bibitem[{{Horns} {et~al.}(2007){Horns}, {Hoffmann}, {Santangelo}, {Aharonian},
  \& {Rowell}}]{hhs+07}
{Horns}, D., {Hoffmann}, A.~I.~D., {Santangelo}, A., {Aharonian}, F.~A., \&
  {Rowell}, G.~P. 2007, A\&A, 469, L17

\bibitem[{{Hotan} {et~al.}(2004){Hotan}, {van Straten}, \&
  {Manchester}}]{hvm04}
{Hotan}, A.~W., {van Straten}, W., \& {Manchester}, R.~N. 2004, Proc. Astr.
  Soc. Aust., 21, 302

\bibitem[{{Johnston} \& {Weisberg}(2006)}]{jw06}
{Johnston}, S., \& {Weisberg}, J.~M. 2006, MNRAS, 368, 1856

\bibitem[{{Kassim} \& {Lazio}(1999)}]{kl99}
{Kassim}, N.~E., \& {Lazio}, T.~J.~W. 1999, ApJ, 527, L101

\bibitem[Kiminki et al.(2007)]{kkk+07}
Kiminki, D.~C., et~al. 2007, \apj, 664, 1102

\bibitem[{{Kn{\"o}dlseder}(2000)}]{kno00}
{Kn{\"o}dlseder}, J. 2000, A\&A, 360, 539

\bibitem[{{Konopelko} {et~al.}(2007){Konopelko}, {Konopelko}, \&
  {Konopelko}}]{kab+07}
{Konopelko}, A., et~al. 2007, ApJ, 658, 1062

\bibitem[{Lorimer \& Kramer(2005)}]{lk05}
Lorimer, D.~R., \& Kramer, M. 2005, Handbook of Pulsar Astronomy (Cambridge
  University Press)

\bibitem[{{Manchester} {et~al.}(2005){Manchester}, {Hobbs}, {Teoh}, \&
  {Hobbs}}]{mhth05}
{Manchester}, R.~N., {Hobbs}, G.~B., {Teoh}, A., \& {Hobbs}, M. 2005, AJ, 129,
  1993

\bibitem[{Manchester {et~al.}(2001)Manchester, Lyne, Camilo, Bell, Kaspi,
  D'Amico, McKay, Crawford, Stairs, Possenti, Morris, \& Sheppard}]{mlc+01}
Manchester, R.~N., et~al. 2001, MNRAS, 328, 17

\bibitem[{{Mart{\'{\i}}} {et~al.}(2007){Mart{\'{\i}}}, {Paredes}, {Ishwara
  Chandra}, \& {Bosch-Ramon}}]{mpib07}
{Mart{\'{\i}}}, J., {Paredes}, J.~M., {Ishwara Chandra}, C.~H., \&
  {Bosch-Ramon}, V. 2007, A\&A, 472, 557

\bibitem[{{Massey} \& {Thompson}(1991)}]{mt91}
{Massey}, P., \& {Thompson}, A.~B. 1991, AJ, 101, 1408 (MT91)

\bibitem[{{Mattana} {et~al.}(2009){Mattana}, {Falanga}, {G{\"o}tz}, {Terrier},
  {Esposito}, {Pellizzoni}, {DeLuca}, {Marandon}, {Goldwurm}, \&
  {Caraveo}}]{mfg+09}
{Mattana}, F., et~al. 2009, ApJ, 694, 12

\bibitem[{{Morgan} {et~al.}(2005){Morgan}, {Byard}, {DePoy}, {Derwent},
  {Kochanek}, {Marshall}, {O'Brien}, \& {Pogge}}]{mbd+05}
{Morgan}, C.~W., {Byard}, P.~L., {DePoy}, D.~L., {Derwent}, M., {Kochanek},
  C.~S., {Marshall}, J.~L., {O'Brien}, T.~P., \& {Pogge}, R.~W. 2005, \aj, 129,
  2504

\bibitem[{{Mukherjee} {et~al.}(2007){Mukherjee}, {Gotthelf}, \&
  {Halpern}}]{mgh07}
{Mukherjee}, R., {Gotthelf}, E.~V., \& {Halpern}, J.~P. 2007, Ap\&SS, 309, 29

\bibitem[{{Mukherjee} {et~al.}(2003){Mukherjee}, {Halpern}, {Gotthelf},
  {Eracleous}, \& {Mirabal}}]{mhg+03}
{Mukherjee}, R., {Halpern}, J.~P., {Gotthelf}, E.~V., {Eracleous}, M., \&
  {Mirabal}, N. 2003, \apj, 589, 487

\bibitem[{{Negueruela} {et~al.}(2008){Negueruela}, {Marco}, {Herrero}, \&
  {Clark}}]{nmhc08}
{Negueruela}, I., {Marco}, A., {Herrero}, A., \& {Clark}, J.~S. 2008, A\&A,
  487, 575

\bibitem[{{Paredes} {et~al.}(2007){Paredes}, {Mart{\'{\i}}}, {Ishwara Chandra},
  \& {Bosch-Ramon}}]{pmib07}
{Paredes}, J.~M., {Mart{\'{\i}}}, J., {Ishwara Chandra}, C.~H., \&
  {Bosch-Ramon}, V. 2007, ApJ, 654, L135

\bibitem[{{Pellizzoni} {et~al.}(2009){Pellizzoni}, {Pilia}, {Possenti},
  {Fornari}, {Caraveo}, {Monte}, {Mereghetti}, {Tavani}, {Argan}, {Trois},
  {Burgay}, {Chen}, {Cognard}, {Costa}, {D'Amico}, {Esposito}, {Evangelista},
  {Feroci}, {Fuschino}, {Giuliani}, {Halpern}, {Hobbs}, {Hotan}, {Johnston},
  {Kramer}, {Longo}, {Manchester}, {Marisaldi}, {Palfreyman}, {Weltevrede},
  {Barbiellini}, {Boffelli}, {Bulgarelli}, {Cattaneo}, {Cocco}, {D'Ammando},
  {DeParis}, {Cocco}, {Donnarumma}, {Fiorini}, {Froysland}, {Galli},
  {Gianotti}, {Harding}, {Labanti}, {Lapshov}, {Lazzarotto}, {Lipari}, {Mauri},
  {Morselli}, {Pacciani}, {Perotti}, {Picozza}, {Prest}, {Pucella},
  {Rapisarda}, {Rappoldi}, {Soffitta}, {Trifoglio}, {Vallazza}, {Vercellone},
  {Vittorini}, {Zambra}, {Zanello}, {Pittori}, {Verrecchia}, {Preger},
  {Santolamazza}, {Giommi}, \& {Salotti}}]{ppp+09}
{Pellizzoni}, A., et al. 2009, \apj, 691, 1618

\bibitem[Predehl \& Schmitt(1995)]{pre95}
Predehl, P., \& Schmitt, J. H. M. M. 1995, A\&A, 293, 889

\bibitem[{Radhakrishnan \& Cooke(1969)}]{rc69a}
Radhakrishnan, V., \& Cooke, D.~J. 1969, Astrophys. Lett., 3, 225

\bibitem[{Ransom(2001)}]{ran01}
Ransom, S.~M. 2001, PhD thesis, Harvard Univ.

\bibitem[{{Ransom} {et~al.}(2002){Ransom}, {Eikenberry}, \&
  {Middleditch}}]{rem02}
{Ransom}, S.~M., {Eikenberry}, S.~S., \& {Middleditch}, J. 2002, AJ, 124, 1788

\bibitem[{{Roberts} {et~al.}(2002){Roberts}, {Hessels}, {Ransom}, {Kaspi},
  {Freire}, {Crawford}, \& {Lorimer}}]{rhr+02}
{Roberts}, M.~S.~E., {Hessels}, J.~W.~T., {Ransom}, S.~M., {Kaspi}, V.~M.,
  {Freire}, P.~C.~C., {Crawford}, F., \& {Lorimer}, D.~R. 2002, ApJ, 577, L19

\bibitem[{Romani \& Yadigaroglu(1995)}]{ry95}
Romani, R.~W., \& Yadigaroglu, I.-A. 1995, ApJ, 438, 314

\bibitem[{Tauris {et~al.}(1994)Tauris, Nicastro, Johnston, Manchester, Bailes,
  Lyne, Glowacki, Lorimer, \& D'Amico}]{tnj+94}
Tauris, T.~M., et~al. 1994, ApJ, 428, L53

\bibitem[{{Thompson}(2001)}]{tho01b}
{Thompson}, D.~J. 2001, in AIP Conf. Ser. 558, High Energy Gamma-Ray
  Astronomy, ed. F.~A. {Aharonian} \& H.~J. {V{\"o}lk} (Berlin: Springer), 103

\bibitem[{{Thompson}(2004)}]{tho04b}
{Thompson}, D.~J. 2004, in Cosmic Gamma-Ray Sources, ed. K.~S. {Cheng} \& G.~E.
  {Romero} (Dordrecht: Kluwer), 149

\bibitem[{Thompson {et~al.}(1999)Thompson, Bailes, Bertsch, Cordes, D'Amico,
  Esposito, Finley, Hartman, Hermsen, Kanbach, Kaspi, Kniffen, Kuiper, Lin,
  Manchester, Matz, Mayer-Hasselwander, Michelson, Nolan, Ogelman, Pohl,
  Ramanamurthy, Sreekumar, Reimer, Taylor, \& Ulmer}]{tbb+99}
Thompson, D.~J., et~al. 1999, ApJ, 516, 297

\bibitem[{{Watters} {et~al.}(2009){Watters}, {Romani}, {Weltevrede}, \&
  {Johnston}}]{wrwj09}
{Watters}, K.~P., {Romani}, R.~W., {Weltevrede}, P., \& {Johnston}, S. 2009,
  \apj, 695, 1289

\bibitem[{Weltevrede \& Johnston(2008)}]{wj08}
Weltevrede, P., \& Johnston, S. 2008, MNRAS, 391, 1210

\bibitem[{{Young} {et~al.}(1999){Young}, {Manchester}, \& {Johnston}}]{ymj99}
{Young}, M.~D., {Manchester}, R.~N., \& {Johnston}, S. 1999, Nature, 400, 848

\end{thebibliography}
\end{document}